%% file: main.tex
\documentclass[twocolumn]{autart}    

\usepackage{graphicx}          

\usepackage{amsmath} 
\usepackage{amssymb}  
\usepackage{cases} 
\usepackage{comment}
\usepackage[normalem]{ulem}
\usepackage{cancel}

\usepackage{tikz}
\usetikzlibrary{positioning}
\usetikzlibrary{shapes,arrows}
\usetikzlibrary{decorations.markings}

\newtheorem{theorem}{Theorem}

\newtheorem{rem}{Remark}

\newcommand{\R}{\ensuremath{\mathbb{R}}}

\begin{document}
\begin{frontmatter}

\title{Fuelling fusion plasmas with pellets: Can neuromorphic control outperform Sigma-Delta modulation?}

\author[Differ,CST]{L.L.T.C. Jansen}\ead{l.l.t.c.jansen@differ.nl},    
\author[CST]{E. Petri}\ead{e.petri@tue.nl},               
\author[Differ,CS]{M. van Berkel}\ead{m.vanberkel@differ.nl},    
\author[CST]{W.P.M.H. Heemels}\ead{m.heemels@tue.nl}  

\address[Differ]{DIFFER - Dutch Institute for Fundamental Energy Research, Eindhoven, The Netherlands}  
\address[CST]{Control Systems Technology (CST) Section, Dept. of Mechanical Eng., Eindhoven University of Technology, The Netherlands}             
\address[CS]{Control Systems (CS) Group, Dept. of Electrical Eng., Eindhoven University of Technology, The Netherlands}


\begin{abstract}                          
Nuclear fusion is a promising clean energy source in which deuterium and tritium fuse inside a magnetically confined plasma in a tokamak, releasing energy. 
A key challenge on the route to practical nuclear fusion is the control of the plasma density which has to be done through adding fuel in the form of deuterium and tritium to the plasma. 
Pellet injection, firing frozen fuel into the plasma, is used to accomplish this. 
Since the injection of a pellet causes an almost instantaneous increase in particle density
compared to the time scales of the plasma dynamics, the problem is of a hybrid nature in which continuous plasma dynamics are interrupted by discrete bursts of particles.
In this paper, we propose a formal hybrid model for this fuelling process and we propose a new, neuron-inspired control method that treats pellets much like spikes as in a brain-like system. 
The neuromorphic controller offers a lightweight solution that naturally fits the hybrid character of pellet fuelling.
For comparison, we also develop a hybrid model of sigma-delta modulation, which is used in current tokamaks. 
For both the neuromorphic controller and the sigma-delta modulation we present formal analysis results for this control problem in nuclear fusion. 
We derive explicit actuator and controller parameter constraints, key for controller tuning, that lead to practical stability guarantees. 
Numerical simulations compare the different controller variants and validate the theoretical results.
\end{abstract}

\end{frontmatter}

\section{Introduction}
Nuclear fusion is a promising and clean energy source in which deuterium and tritium (isotopes of hydrogen) fuse inside a magnetically confined plasma in a tokamak, releasing energy.
Maximizing the energy output of a tokamak requires precise plasma density control by fuelling the plasma with deuterium and tritium: 
higher particle density increases fusion rates but exceeding safety limits can trigger disruptive plasma instabilities.
Pellet injection, i.e., firing frozen deuterium-tritium pellets into the plasma, is the most effective fuelling method, especially for future high-field reactors where gas injection is inefficient due to edge transport barriers \cite{Geulin:2022}, \cite{Romanelli:2015}.
The use of pellets for plasma density control presents unique control challenges. 
The first challenge is related to specific actuator characteristics, 
with existing pellet injection actuators falling into two main categories:
\begin{itemize}
    \item \textit{Gas gun actuators} accelerate pellets using compressed gas.
    They are used in, e.g., the Joint European Torus (JET) \cite{Combs-gasgun} and the stellarator Wendelstein 7-X \cite{W7X-gasgun}. 
    Their firing rate is limited by a pellet preparation time, resulting in a non-negligible delay between successive launches.
    \item \textit{Centrifuge actuators} accelerate pellets on a rotating disk via centrifugal force,
    and are used in, e.g.,  the Axially Symmetric Divertor Experiment Upgrade (AUG) \cite{Ploeckl-centrifuge}.
    Pellets can only be launched when aligned with the ejection port, requiring synchronization and allowing launches only at discrete, periodic times. 
    A pellet preparation time also applies.
\end{itemize}
Hence, for both type of actuators a pellet preparation time is present, and additionally, the centrifuge-type actuator allows only launches at periodic times.

A second challenge for plasma density control is that we need to deal with the discrete nature of the pellets.
When a frozen deuterium-tritium pellet is injected, it ablates and ionizes over 
1-2~ms, which can be considered as
a single discrete event,
compared to the time scales of other dynamical processes in the plasma,
delivering particles in a short burst rather than in a continuous flux. 
This results in actuation spikes and (almost) discontinuous density trajectories, which transforms density regulation into a hybrid control problem \cite{Goebel:hybrid}, \cite{Heemels:2010} with continuous dynamics to describe the plasma evolution with losses from transport, plasma-wall interactions, pumping, etc., and discrete events to describe the instantaneous pellet-induced density jumps.

Although many tokamaks employ pellet injectors, advanced control strategies remain limited. 
Most approaches model pellet fuelling as a continuous input, e.g., \cite{Ravensbergen-control}, \cite{pellets-JET}, \cite{Lang-controller}, \cite{Kudlacek-overview}, enabling standard linear control design while neglecting the discrete nature of pellet injection.
At AUG, a PI feedback controller generates a continuous particle flux signal that is converted into discrete injections via sigma-delta modulation \cite{Lang-controller}, \cite{Ploeckl-SDM}. 
Similarly, JET treats pellet injectors as equivalent continuous actuators within a PID feedback-feedforward framework, with filtering mitigating discrete effects \cite{pellets-JET}. 
While effective for current devices, such approximations are insufficient for next-generation reactors such as ITER and DEMO \cite{discrete-pellets1}, \cite{discrete-pellets2}.
Model predictive control with discrete decisions has been proposed \cite{Orrico-MPC1}, but real-time implementation can be challenging. 
This motivates the development of controllers that account for the discrete nature of pellet injection while remaining real-time feasible and performant for large tokamaks.

This work explores two closely related control schemes: neuromorphic (NM) control, see, e.g.,
\cite{Petri}, \cite{Ribar}, \cite{Medvedeva}, \cite{Petri25}, and sigma-delta modulation (SDM) \cite{Inose}, \cite{Reiss}.
Both controllers have a computationally lightweight implementation and can be used in a single-input single-output feedback loop as is illustrated in Fig.~\ref{fig:block_diagram}.
SDM is inspired by the current pellet control scheme on AUG as described in \cite{Ploeckl-SDM}.
The NM controller is proposed in this paper for pellet fuelling, based on and integrate-and-fire neuron model as in \cite{Petri}.
In both the NM and SDM case, the controller integrates the error between the reference and measured density through an internal state.
When the internal integrator reaches a positive threshold, the controller emits a spike corresponding to a pellet injection. 
Pellet injections are constrained to discrete time instants imposed by the injector hardware, potentially introducing a delay between the threshold crossing and the pellet delivery.
When a pellet is fired, the integrator is reset, and here lies the main difference between SDM and NM control.
In SDM, the threshold value is subtracted from the integrator, while in the NM controller, the state is reset to 0 after firing.

\tikzstyle{block} = [draw, rectangle, 
    minimum height=3em, minimum width=2em]
\tikzstyle{sum} = [draw, circle, node distance=1cm]
\tikzstyle{input} = [coordinate]
\tikzstyle{output} = [coordinate]
\tikzstyle{pinstyle} = [pin edge={to-,thin,black}]
\tikzstyle{vec->}=[line width=1mm,-stealth,postaction={draw, line width=1mm, shorten >=4mm, -}]
\tikzset{->-/.style={decoration={
  markings,
  mark=at position #1 with {\arrow{>}}},postaction={decorate}}}

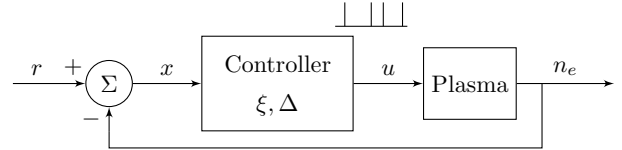
\begin{figure}[t!]
    \centering
\begin{tikzpicture}[auto, node distance=4em, scale=0.9, every node/.style={scale=0.9},>=latex']
    \node [input](input){$r$};
    \node [sum, right of=input, node distance=4em](sum){$\Sigma$};
    \node [block, right of=sum, minimum width=2em, node distance=7em](neuron){\begin{tabular}{c} Controller \\ $\xi, \Delta$ \end{tabular}};
    \node [block, right of=neuron, minimum width=3em, node distance=8em](plant){Plasma};
    \node [input, right of=plant, node distance=3em](tmp1){};
    \node [input, below of=sum, node distance=2em](min){};
    \node [input, left of=sum, node distance=2em](plus){};
    \node [output, minimum width=2em, node distance=3em, right of=tmp1](output){$n_e$};
    \node [output, minimum width=2em, node distance=2.7em, below of=tmp1](tmp2){};

    \node [input, right of=neuron, above of=neuron, node distance=2.4em](spikea){};
    \node [input, right of=spikea, node distance=3em](spikeb){};
    \node [input, right of=spikea, node distance=0.4em](spike11){};
    \node [input, above of=spike11, node distance=1em](spike12){};
    \node [input, right of=spikea, node distance=1.5em](spike21){};
    \node [input, above of=spike21, node distance=1em](spike22){};
    \node [input, right of=spikea, node distance=2em](spike31){};
    \node [input, above of=spike31, node distance=1em](spike32){};
    \node [input, right of=spikea, node distance=2.8em](spike41){};
    \node [input, above of=spike41, node distance=1em](spike42){};
    \draw [-]  (spikea) -- (spikeb);
    \draw [-]  (spike11) -- (spike12);
    \draw [-]  (spike21) -- (spike22);
    \draw [-]  (spike31) -- (spike32);
    \draw [-]  (spike41) -- (spike42);

    \draw [draw,-] node [near end] {} (input) -- node {$r$}(plus);
    \draw [->] (neuron) -- node {$u$} (plant);
    \draw [->] (plant) -- node{$n_e$} (output);
    \draw [->] (min) -- node{$-$} (sum);
    \draw [->] (sum) -- node {$x$} (neuron);
    \draw [->] (plus) -- node {$+$} (sum);
    \draw [-] (tmp1) -- (tmp2) -| (min);

\end{tikzpicture}
  \caption{System with spiking controller (NM or SDM), reference input $r$, plasma density $n_e$, error $x$ and control action $u$. 
  The controller, either NM or SDM, is defined by an internal integrator $\xi$ and a threshold value $\Delta$. 
}
\label{fig:block_diagram}
\end{figure}

Next to proposing the NM controller, and providing a hybrid modelling setup for the closed-loop NM and SDM systems, this work provides a rigorous theoretical analysis of pellet-based plasma density control, with formal stability guarantees for both NM control and SDM. 
The main contribution is the design of a NM controller for pellet fuelling with a complete proof of its principal stability properties, along with proofs of all derived propositions, addressing a gap in the existing literature that emphasises practical implementation over formal analysis. 
For both controllers, convergence to the reference density is proven up to a bounded error. 
Finally, numerical case studies compare all controller variants under identical conditions, validating the theoretical results and assessing performance, including an evaluation of SDM integrator wind-up and two mitigation strategies.

The paper is structured as follows.
Mathematical preliminaries are given in Section~\ref{section:preliminaries}. 
In Section~\ref{section:model} we present the hybrid model for plasma density with pellet injection.
Section~\ref{section:NM} describes the NM controller using the hybrid model, and states the main stability theorem of this paper. 
This is followed by an analysis of the SDM in Section~\ref{section:SDM}, 
discussing integrator wind-up and proposing two mitigation methods.
In Section~\ref{section:refr} both NM and SDM models are extended with a pellet preparation time.
Finally, Section~\ref{section:conclusion} outlines the conclusions and presents future work directions.

\section{Preliminaries}
\label{section:preliminaries}
The notation $\mathbb{R}$ stands for the set of real numbers, 
$\mathbb{R}_{\geqslant 0} := [0,+\infty)$ and
$\mathbb{R}_{> 0} := (0,+\infty)$.
We use $\mathbb{Z}$ to denote the set of integers, 
$\mathbb{Z}_{\geqslant 0} := \{0, 1, 2, \ldots \}$ and 
$\mathbb{Z}_{> 0} := \{1, 2, \ldots \}$.
We present the NM controller in a closed loop with the plant, following the hybrid system in \cite{Petri}, with the formalism in \cite{Goebel:hybrid},
\begin{subnumcases}{\mathcal{H} : \label{eq:hybrid_system_general}}
        \hfill \dot{q} \; = \; F(q),  & $q \in  \mathcal{C}$ \label{eq:hybrid_system_general_flow}\\
        q^+ \; \in \; G(q),  & $q \in  \mathcal{D}$,\label{eq:hybrid_system_general_jump}
\end{subnumcases}
where $\mathcal{C} \subseteq \mathbb{R}^{n_q}$ is the flow set,
$\mathcal{D} \subseteq \mathbb{R}^{n_q}$ is the jump set,
$F : \mathbb{R}^{n_q} \to \mathbb{R}^{n_q}$ is the flow map, and 
$G : \mathbb{R}^{n_q} \rightrightarrows \mathbb{R}^{n_q}$ is the set-valued jump map,
with $n_q \in Z_{\geqslant0}$ the dimension of the state vector $q$.
We consider hybrid time domains as defined in \cite[Definition~2.3]{Goebel:hybrid} and we use the notion of solutions for system \eqref{eq:hybrid_system_general} as in \cite[Definition 2.6]{Goebel:hybrid}. 
Given a solution $q$ for system \eqref{eq:hybrid_system_general} and its hybrid time domain dom $q$, we define 
$\sup_t  \text{dom } q:= \sup\{t \in \mathbb{R}_{\geqslant 0} : \exists \;j \in \mathbb{Z}_{\geqslant 0} \text{ such that } (t, j) \in \text{dom } q\} $ and
$\sup_j \text{dom } q:= \sup\{j \in \mathbb{Z}_{\geqslant 0} : \exists \; t \in \mathbb{R}_{\geqslant 0} \text{ such that } (t, j) \in \text{dom } q\}$. 
The notation $(t,j) \geqslant (t^\star, j^{\star})$ means that $t \geqslant t^\star$ and $j \geqslant  j^{\star}$,  where $(t,j),(t^\star,j^\star)\in\R_{\geqslant 0}\times\mathbb{Z}_{\geqslant0}$. 
Moreover, we say that the solution $q$ is complete if its hybrid time domain is unbounded and $t$-complete if $\sup_t \text{dom } q = +\infty$. 
The tangent cone to the set $S \subset \R^{n_q}$ with $n_q \in Z_{\geqslant0}$ at the point $q \in \R^{n_q}$
is denoted $T_S(q)$, see \cite[Definition~5.12]{Goebel:hybrid}.
We say that system \eqref{eq:hybrid_system_general} satisfies the hybrid basic conditions \cite[Assumption~6.5]{Goebel:hybrid} if \textit{(i)} $F$ is continuous and $C \subset \textnormal{dom } F$; \textit{(ii)} $G$ is outer semicontinuous and locally bounded relative to $D$ and $D \subset \textnormal{dom } G$; \textit{(iii)} $\mathcal{C}$ and $\mathcal{D}$ are closed, 
see for further details \cite{Goebel:hybrid}.

\section{Hybrid model for plasma density with pellet injection}
\label{section:model}
Building on the hybrid systems formalism introduced in \eqref{eq:hybrid_system_general}, we model the plasma density by splitting the system into two dynamical behaviours: \textit{continuous flow} dynamics \eqref{eq:hybrid_system_general_flow} that capture how the system evolves between control actions, and \textit{discrete jumps} \eqref{eq:hybrid_system_general_jump} that represent pellet injections. 
We adopt a reduced-order model for the plasma density dynamics based on the particle inventory model of \cite{Blanken:2018}, assuming instantaneous ionization, recombination, and plasma equilibration relative to gas injection and particle confinement timescales. 
Neglecting actuator and sensor delays leads to a first-order electron density model during flow,
\begin{equation}
    \dot{n}_e = -\frac{1}{\tau}n_e,
    \label{eq:openloop_density}
\end{equation}
where $n_e(t) \in \mathbb{R}_{\geqslant 0}$ represents the electron density in the plasma at time $t \in \mathbb{R}_{\geqslant 0}$, and $\tau \in \mathbb{R}_{> 0}$ a time constant.
In \cite{Derks}, results from system identification experiments on MAST-U, show a linear core density response in the 7-40 Hz range, 
successfully fitting a first-order plus time delay model to the experimental data.
While this representation oversimplifies the complex plasma physics, it retains the most relevant dynamics essential for controller design.
To regulate the plasma density to a constant reference value $r \in \mathbb{R}_{\geqslant 0}$,
we define the error between the desired reference value and the measured density as $x(t) := r - n_e(t) \in \mathbb{R}$, $t \in \mathbb{R}_{\geqslant 0}$.
We get from \eqref{eq:openloop_density} that during flow, $x(t)$ evolves according to
\begin{equation}
    \dot{x} = \frac{1}{\tau} (r - x).
    \label{eq:openloop_x}
\end{equation}
Physical constraints impose a fundamental bound: because the electron density $n_e$ cannot be negative, the error $x$ is upper-bounded by $r$.
Fig.~\ref{fig:block_diagram} illustrates the closed-loop system with the controller and the plasma.

To fuel the plasma, pellet injection is modelled as an instantaneous density increase.
This is represented by a Dirac delta pulse input $m_p \delta(t-t_j)$, 
where spike amplitude $m_p \in \R_{\geqslant 0}$ is the number of particles per pellet, with time $t \in \R_{\geqslant 0}$, spiking time $t_j$ and $j \in \mathbb{Z}_{> 0}$.
The resulting plasma density increase is defined as $\alpha:=\frac{m_p}{V}  \in \R_{\geqslant 0}$, where $V$ is the plasma volume, assuming no losses.
Since the particle deposition of a pellet into the plasma happens fast, 
relative to the sensor timescale, 
it is justified to model it as a single discrete event in the form of \eqref{eq:hybrid_system_general_jump}, which becomes $n_e^+=n_e+\alpha$,
and thus for the error
\begin{align}
    x^+ & = x - \alpha.
    \label{eq:jump_x}
\end{align}
Having both the models for the continuous flow \eqref{eq:openloop_x} and the discrete jumps \eqref{eq:jump_x}, we can now present controller designs for this system. 

\section{Neuromorphic controller}
\label{section:NM}
We first introduce a controller inspired by the work of \cite{Petri} that is based on a spiking integrate-and-fire neuron model, as modelled in \cite{Lapicque}, \cite{Gerstner}. 
In this type of controller, an accumulator variable $\xi \in \R_{\geqslant 0}$, that represents the neuronal membrane potential, integrates the neuron's input value until reaching a threshold $\Delta \in \R_{> 0}$.
Upon threshold crossing, it spikes and $\xi$ resets to~zero.
Extending the model in \cite{Petri}, we provide a novel NM controller that accounts for the pellet actuator behaviour through three key ingredients, as introduced in a preliminary conference version of this work, presented in \cite{Jansen}.
\begin{itemize}
    \item The single-neuron controller produces only positive outputs, reflecting the physical constraint that pellets can only be added to (not removed from) the plasma.
    \item The controller targets a specified reference density rather than an equilibrium at zero in \cite{Petri}, requiring stability of the error between desired and actual plasma densities.
    \item Our controller restricts firing events to discrete, periodic intervals synchronised with the pellet actuator's operational cycle.
    This design accommodates centrifuge type actuators, though it can easily be extended to gas gun systems, which is briefly discussed in Section \ref{section:refr}.
    This is distinctly different from the design in \cite{Petri}, which allowed firing of spikes at all times without restriction.
\end{itemize}

The NM controller has the following hybrid dynamics. 
The neuron membrane potential $\xi$ evolves according to 
\begin{equation}
    \dot{\xi} = \max(0,x) = 
    \begin{cases}
        0  & \text{if} \quad    x \leqslant 0 \\
        x & \text{if} \quad     x > 0,
    \end{cases}
    \label{eq:neuron_flow}
\end{equation}
between spikes, and
\begin{equation}
    \xi^+ = 0 ,
    \label{eq:reset_neuron}
\end{equation}
at jumps, i.e., the neuron membrane resets to $0$.
A jump happens when
\begin{equation}
    \xi \geqslant \Delta
    \label{eq:jump_cond_neuron}
\end{equation}
is satisfied, and a pellet launch is possible.
In this paper, we consider a centrifuge-type actuator \cite{Ploeckl-centrifuge}
for pellet fuelling, which requires synchronisation between the controller and moment of injection.
This required synchronisation causes that pellet launches can only occur at multiples of periodic time $T_c \in \R_{\geqslant0}$, which is the time it takes the centrifuge to make a full rotation, see \cite{Ploeckl-centrifuge}.
We can model this by using a hybrid timer with continuous flow dynamics
\begin{equation}
    \dot{T} = 1,
    \label{eq:timer_flow}
\end{equation}
where $T \in \mathbb{R}_{\geqslant 0}$.
When a launch slot is available at
\begin{equation}
    T = T_c ,
    \label{eq:jump_cond_timer}
\end{equation}
the timer resets to zero, with discrete jump dynamics
\begin{equation}
    T^+ = 0 .
    \label{eq:reset_timer}
\end{equation}
Hence, a jump according to \eqref{eq:jump_x} occurs when $T=T_c$ and $\xi \geqslant\Delta$.
Note that this timer synchronization problem is not present in \cite{Petri}, where neurons generate spikes as soon as they reach the threshold $\Delta$. 

\section{Model and analysis with neuromorphic controller}
\label{section:analysisNM}
\subsection{Hybrid model}
To combine all of the above, 
we define the overall state of the hybrid system as $q := (x,\xi,T) \in \mathcal{X}$ with
$\mathcal{X}:=\{(x,\xi,T) \in \mathbb{R}^3 : (x \leqslant r) \wedge (\xi \geqslant 0) \wedge (T \geqslant 0) \}$, with $r \in \R_{\geq 0}$ the desired reference, and we obtain the hybrid system \eqref{eq:hybrid_system_general},
where the flow map $F$ is defined, for any $q \in \mathcal{C}$, from \eqref{eq:openloop_x}, \eqref{eq:neuron_flow} and \eqref{eq:timer_flow},
\begin{equation}
    F(q) := 
    \begin{pmatrix}
        -\frac{1}{\tau}(x - r) \\
        \max(0,x) \\
        1
    \end{pmatrix}.
    \label{eq:flowmap}
\end{equation}
The flow set $\mathcal{C}$ in \eqref{eq:hybrid_system_general} is defined as
\begin{equation}
    \mathcal{C} := \{ q \in \mathcal{X}: (T \leqslant T_c) \}.
    \label{eq:flowSet}
\end{equation}
The jump set $\mathcal{D}$ in \eqref{eq:hybrid_system_general} is defined as, from \eqref{eq:jump_cond_timer},
\begin{equation}
    \mathcal{D} := \{ q \in \mathcal{X} : T \geqslant T_c \}. \label{eq:jumpset}
\end{equation}
Note that the way $\mathcal{C}$ and $\mathcal{D}$ are defined jumps take place when $T = T_c$, i.e., \eqref{eq:jump_cond_timer}.
To consider the two possible jumps, i.e., only a timer reset as in \eqref{eq:reset_timer}, or a timer reset with neuron spike as in \eqref{eq:jump_x}, \eqref{eq:reset_neuron} and \eqref{eq:reset_timer}, we write $\mathcal{D}$ as $\mathcal{D} := \mathcal{D}_1 \cup \mathcal{D}_2$, with 
\begin{align}
    \mathcal{D}_1 &:= \{ q \in \mathcal{X}: (T \geqslant T_c ) \wedge (\xi \leqslant \Delta )\} \label{eq:jumpsetD1}\\
    \mathcal{D}_2 &:= \{ q \in \mathcal{X}: (T \geqslant T_c ) \wedge ( \xi \geqslant \Delta )\}.
    \label{eq:jumpsetD2}
\end{align}
The jump map $G$ in \eqref{eq:hybrid_system_general} is defined for any $q \in \mathcal{D}$, from \eqref{eq:jump_x}, \eqref{eq:reset_neuron} and \eqref{eq:reset_timer}, as
\begin{equation}
    G(q) := G_1(q) \cup G_2(q)
    \label{eq:jump_map}
\end{equation}
with
\begin{equation}
    G_{1}(q) := 
    \begin{cases}
    \left\{
        \begin{pmatrix}
            x \\
            \xi \\
            0
        \end{pmatrix} 
        \right\}
        & q \in \mathcal{D}_1 \\
        \hfill \emptyset \hfill  & q \notin \mathcal{D}_1,
    \end{cases}
    \label{eq:jump_map_G1}
\end{equation}
corresponding to no pellet launch, and
\begin{equation}
    G_{2}(q) := 
    \begin{cases}
    \left\{
        \begin{pmatrix}
            x - \alpha \\
            0 \\
            0
        \end{pmatrix} 
        \right\}
        & q \in \mathcal{D}_2 \\
        \hfill \emptyset \hfill  & q \notin \mathcal{D}_2,
    \end{cases}
    \label{eq:jump_map_G2}
\end{equation}
related to a pellet launch.

\subsection{Main analysis}
In this section we present our main analysis result.
\begin{prop}
\label{prop:hybrid-model-basic-properties} 
The following properties hold for system \eqref{eq:hybrid_system_general} with \eqref{eq:flowmap}-\eqref{eq:jump_map_G2}.
\begin{enumerate}
\item[(i)] The hybrid basic conditions are satisfied.
\item[(ii)] Zeno behaviour cannot occur.
\item[(iii)] Maximal solutions are $t$-complete.
\end{enumerate}

\end{prop}The proof of Proposition \ref{prop:hybrid-model-basic-properties} is given in Appendix \ref{app:basic_properties}.
In the next theorem we establish a practical stability property for closed-loop system \eqref{eq:hybrid_system_general} with the NM controller in \eqref{eq:flowmap}-\eqref{eq:jump_map_G2}.

\begin{theorem}
\label{theorem:main}
Consider system \eqref{eq:hybrid_system_general}, with $F,\mathcal{C},G,\mathcal{D}$ as in \eqref{eq:flowmap}-\eqref{eq:jump_map_G2}, 
with $x(0,0) \leqslant r$ and $\xi(0,0)=T(0,0)=0$,
with a desired reference value $r \in \mathbb{R}_{>0}$, jump size $\alpha \in \mathbb{R}_{>0}$, and $r>\alpha$.
For all 
\begin{equation}
    T_c \in \left(0,\tau \ln{\frac{r}{r-\alpha}}\right],    
    \label{eq:Tc_condition}
\end{equation}
select 
\begin{align}
\Delta \in \left(0,r \tau \ln{\frac{r}{r-\alpha}} - r \tau \left(1- \frac{r-\alpha}{r} e^{\frac{T_c}{\tau}}\right) - r T_c\right]\text{}
\label{eq:Delta_condition}
\hspace*{19pt}
\raisetag{1.4\baselineskip} 
\end{align}
and define $\tau_d := \tau\ln{(\frac{r}{r-\alpha})} \in \R_{>0}$. 
Then, any solution $q$ of system \eqref{eq:hybrid_system_general} satisfies for all $(t,j) \in \textnormal{dom }q$,
\begin{gather}
\begin{aligned}
 -\alpha < x(t,j) \leqslant \gamma^{\left(\frac{t}{\tau_d} -1\right)}  x(0,0) + \alpha , &\\
   \text{if } \;  x(0,0) &>0 \\
   \min\left(r-e^{-\frac{t}{\tau}}\left(r-x(0,0)\right),-\alpha\right) < x(t,j) \leqslant  \alpha , &\\
   \text{if } \;  x(0,0) &\leqslant 0 ,
\end{aligned}
\label{eq:ultimate_bound}
\hspace*{16pt}
\raisetag{2.2\baselineskip} 
\end{gather}
with $\gamma := \frac{r-\alpha}{r} \in (0,1)$.
Also, 
$\lim\limits_{t+j \to \infty} \sup |x(t,j)| \leqslant \alpha$.
\end{theorem}

Theorem \ref{theorem:main}, whose proof is given in Appendix \ref{app:main}, establishes conditions on $\Delta$ and $T_c$, under which the neuromorphic controller guarantees practical stability for system \eqref{eq:hybrid_system_general}. 
Satisfying the conditions in \eqref{eq:Tc_condition} and \eqref{eq:Delta_condition}, guarantees convergence of the density error $x$ to a neighbourhood $(-\alpha, \alpha]$ around the origin.

If we want an ultimate bound of size $\alpha$,
condition \eqref{eq:Tc_condition} specifies the maximum allowable rotation period $T_c$ for the centrifuge actuator, which equivalently defines its minimum required speed. 
This constraint depends on three key parameters:
\begin{description}
    \item[Pellet size ($\alpha$)] Larger pellets (increased $m_p$, and $\alpha = \frac{m_p}{V}$) permit slower actuator rotation (larger $T_c$) while still being able to reach a certain reference density, as each injection delivers more particles.
    However, the larger the pellet size, the larger is the ultimate bound of the convergence.
    \item[Reference density ($r$)] Higher target densities require faster rotation (smaller $T_c$) to maintain sufficient fuelling rates.
    \item[Confinement time ($\tau$)] Shorter particle confinement times necessitate faster actuation (smaller $T_c$) to compensate for rapid density decay between injections.
\end{description}
While \eqref{eq:Tc_condition} can guide new reactor design by specifying centrifuge requirements, for operational tokamaks it conversely determines the maximum density achievable by current pellet systems,
i.e., given $\alpha$, $\tau$, and $T_c$, 
as
\begin{equation}
    r_\textnormal{max} = \frac{e^{\frac{T_c}{\tau}}}{e^{\frac{T_c}{\tau}}-1} \alpha \in \R_{\geqslant 0}.
    \label{eq:r_max}
\end{equation}
Condition \eqref{eq:Delta_condition} establishes an allowable range for tuning the neuron threshold parameter $\Delta$.
The threshold directly influences the controller's response characteristics, where a small $\Delta$ leads to fast response to an error ($\xi$ will reach $\Delta$ sooner), while a larger $\Delta$ allows for a longer time before launching the next pellet.
Note that this effect mainly affects steady-state performance, 
where it may move within the limits of the ultimate bound established in Theorem~\ref{theorem:main},
and not transient response where the error is typically large enough to trigger a fast response.
The effect is illustrated in the next example.

\begin{figure}
    \begin{center}
        \includegraphics[width=0.47\textwidth]{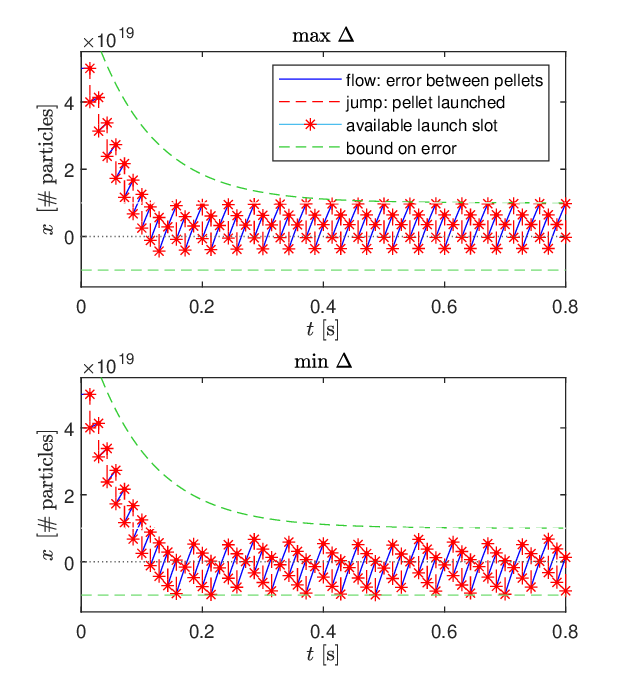} 
        \caption{Numerical simulation of the neurmorphic controller, with $\alpha =1 \cdot 10^{19}$ particles/m$^3$, $r = 5 \cdot 10^{19}$~particles/m$^3$, $\tau = 0.1$~s, $T_c=0.0143$~s $\in (0, 0.0223]$~s from condition \eqref{eq:Tc_condition} of Theorem~\ref{theorem:main}, and $x(0,0)=r$.
        In the top figure, with $\Delta = 1.569 \cdot 10^{16}$, the error is at the upper limit of the bound.
        In the bottom figure, with $\Delta =1$, the error is at the lower limit of the bound.
        } 
        \label{fig:delta_min_max}
    \end{center}
\end{figure}

A numerical simulation of the closed-loop system in Fig.~\ref{fig:delta_min_max} shows the impact of tuning $\Delta$
for a tokamak with a volume of 13~m$^3$ with pellets consisting of $1.3 \cdot 10^{20}$~particles. 
A pellet increases the plasma electron density by $\alpha =1 \cdot 10^{19}$ particles/m$^3$. 
The desired reference value $r = 5 \cdot 10^{19}$~particles/m$^3$ and $\tau = 0.1$~s.
We take $n_e(0)=0$ as initial condition, hence the initial error is $x=r-n_e=r$.
The pellet actuator is a centrifuge working at 70~Hz, resulting in $T_c=0.0143$~s,
which is contained in the range
$(0, 0.0223]$ obtained from condition \eqref{eq:Tc_condition} of Theorem~\ref{theorem:main}.
We can choose $\Delta \in (0,1.569 \cdot 10^{16}]$.
The top figure shows the results for $\Delta = 1.569 \cdot 10^{16}$, and the bottom figure shows the results for $\Delta =1$.
When the system converges towards its ultimate bound, a small $\Delta$ pushes the error to the bottom of this range, and a larger $\Delta$ permits the error to remain near the upper limit instead.
Note that in this case, choosing $\Delta$ in between, can result in a smaller ultimate bound, however, this is not a guarantee and depends on the parameters.

This behaviour can change significantly based on operating conditions.
At maximum density targets, $r \to r_{max}$ in \eqref{eq:r_max}, all launch slots must be used, leaving no room to tune $\Delta$, as we get 
the upperbound for $\Delta$ in \eqref{eq:Delta_condition} almost equal to 0.
At lower density targets, the impact of tuning $\Delta$ becomes visible as in Fig. \ref{fig:delta_min_max}.

The condition $r > \alpha$, where $\alpha$ quantifies the pellet's density contribution, arises naturally from the system's physical constraints.
In scenarios where $r \leqslant \alpha$, pellet injection fails to provide meaningful control benefits.
Indeed, the open-loop system \eqref{eq:openloop_x} is asymptotically stable with $x$ converging to $r$ over time without any pellet actuation.
So a reference $r \leqslant \alpha$ is not meaningful.

\begin{rem}
Note that the initial condition for the neuron accumulator variable $\xi(0,0)=0$ and the timer $T(0,0)=0$ are design choices to simplify the proof of Theorem \ref{theorem:main}.
Any other choice for the initial conditions would not compromise the stability result.
As the system is inherently stable, a simple translation in the hybrid time $(t,j)$, making the first jump in $\xi$ the new starting point, $(t_1,1) \rightarrow (0,0)$, will result in a system in which the conditions, and thus the result, from Theorem \ref{theorem:main} still holds.
\end{rem}

\section{Sigma-delta modulation}
\label{section:SDM}
We now examine SDM, a control strategy close to the new NM controller presented above, but with modified reset conditions.
Originally developed for analog-to-digital conversion \cite{Inose}, \cite{Reiss}, the ability to encode continuous signals into discrete pulses makes SDM naturally suited for converting a desired particle flux into timed pellet injections.
A control scheme based on first-order SDM has been successfully implemented in AUG's pellet injection system \cite{Ploeckl-SDM}, albeit without any formal analysis.
Here we will provide such a formal analysis for the first time.

\subsection{Hybrid model}
\label{subsection:hybridmodel_SDM}
\begin{figure}
    \begin{center}
        \includegraphics[width=0.47\textwidth]{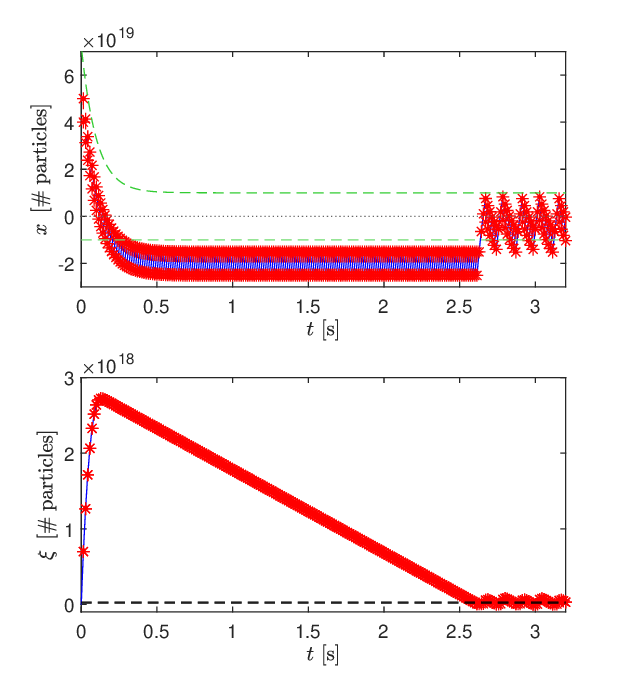} 
        \caption{Numerical simulation of SDM using the same values as in Fig. \ref{fig:delta_min_max} ($\alpha =1 \cdot 10^{19}$ particles/m$^3$, $r = 5 \cdot 10^{19}$~particles/m$^3$, $\tau = 0.1$~s, $T_c=0.0143$~s, $\Delta = 1.569 \cdot 10^{16}$, $x(0,0)=r$).
        The error shows clear undershoot in the transient phase.
        Once the integrator $\xi$ has been emptied, the error establishes a limit cycle with the same ultimate upper bound as the neuromorphic controller, but continues to undershoot the lower bound.} 
        \label{fig:overshoot}
    \end{center}
\end{figure}

To model the SDM controller, we implement a revised reset condition in the hybrid model of Section~\ref{section:NM}.
We use
\begin{equation}
    \xi^+ = \xi - \Delta,
    \label{eq:reset_neuron_SDM}
\end{equation}
instead of the reset in \eqref{eq:reset_neuron}, 
where we recall that $\Delta \in \R_{>0}$ is the threshold from \eqref{eq:jump_cond_neuron}.
This leads to the jump map
\begin{equation}
    G_{\text{SDM}}(q) := G_{\text{SDM}1}(q) \cup G_{\text{SDM}2}(q)
    \label{eq:jump_map_sigmadelta}
\end{equation}
with
\begin{equation}
    G_{\text{SDM}1}(q) := 
    \begin{cases}
    \left\{
        \begin{pmatrix}
            x \\
            \xi \\
            0
        \end{pmatrix} 
        \right\}
        & q \in \mathcal{D}_1 \\
        \hfill \emptyset \hfill  & q \notin \mathcal{D}_1
    \end{cases}
    \label{eq:jump_map_G1_sigmadelta}
\end{equation}
\begin{equation}
    G_{\text{SDM}2}(q) := 
    \begin{cases}
    \left\{
        \begin{pmatrix}
            x - \alpha \\
            \xi - \Delta \\
            0
        \end{pmatrix} 
        \right\}
        & q \in \mathcal{D}_2 \\
        \hfill \emptyset \hfill  & q \notin \mathcal{D}_2,
    \end{cases}
    \label{eq:jump_map_G2_sigmadelta}
\end{equation}
where $\mathcal{D}_1$ and $\mathcal{D}_2$ are defined in the same way as for the NM controller in \eqref{eq:jumpsetD1}-\eqref{eq:jumpsetD2}.
While this modification appears minor, the significant order-of-magnitude difference between $\xi$ and $\Delta$ creates undershoot of the error.
This effect, which is comparable to integrator wind-up, is shown in a numerical simulation in Fig. \ref{fig:overshoot}, using the same values as the NM controller in Section~\ref{section:analysisNM}.
This behaviour occurs when $\xi$ increases more than $\Delta$ before a next pellet can be launched.
Since each pellet launch removes only $\Delta$ from $\xi$, it requires multiple pellet launches to get $\xi$ below $\Delta$ again.
Undershoot in error ($x<0$) equals an overshoot in electron density ($n_e>r$), 
and, contrary to the NM controller, we cannot guarantee any bounds, as the the controller will use all available launch slots until $\xi \in [0, \Delta)$.
Consequently, the controller drives the density to the actuator's operational limits, determined by $T_c$ and $\alpha$, and ignoring $r$.
This can cause a plasma disruption and must be prevented.
There are several ways to solve this issue, in the next sections we further investigate input clipping and adjusting the reset condition.

\subsection{Input clipping}
\label{subsec:ic}
\begin{figure}
    \centering
\begin{tikzpicture}[auto, node distance=4em, scale=0.75, every node/.style={scale=0.75},>=latex']
    \node [input](input){$r$};
    \node [sum, right of=input, node distance=4em](sum){$\Sigma$};
    \node [block,minimum size=1cm, right of=sum,node distance=5em] (sat) {};
    \node [block, right of=sat, minimum width=2em, node distance=8em](neuron){Controller};
    \node [block, right of=neuron, minimum width=3em, node distance=8em](plant){Plasma};
    \node [input, right of=plant, node distance=3em](tmp1){};
    \node [input, below of=sum, node distance=2em](min){};
    \node [input, left of=sum, node distance=2em](plus){};
    \node [output, minimum width=2em, node distance=3em, right of=tmp1](output){$n_e$};
    \node [output, minimum width=2em, node distance=4em, below of=tmp1](tmp2){};

    \node [input, right of=neuron, above of=neuron, node distance=2em](spikea){};
    \node [input, right of=spikea, node distance=3em](spikeb){};
    \node [input, right of=spikea, node distance=0.4em](spike11){};
    \node [input, above of=spike11, node distance=1em](spike12){};
    \node [input, right of=spikea, node distance=1.5em](spike21){};
    \node [input, above of=spike21, node distance=1em](spike22){};
    \node [input, right of=spikea, node distance=2em](spike31){};
    \node [input, above of=spike31, node distance=1em](spike32){};
    \node [input, right of=spikea, node distance=2.8em](spike41){};
    \node [input, above of=spike41, node distance=1em](spike42){};
    \draw [-]  (spikea) -- (spikeb);
    \draw [-]  (spike11) -- (spike12);
    \draw [-]  (spike21) -- (spike22);
    \draw [-]  (spike31) -- (spike32);
    \draw [-]  (spike41) -- (spike42);

    \draw[very thick]   (sat.center) -- ++ (-1mm,-3mm) -- ++ (-2mm,0)
                        (sat.center) -- ++ (+1mm,+3mm) -- ++ (+2mm,0);

    \draw [draw,-] node [near end] {} (input) -- node {$r$}(plus);
    \draw [->] (neuron) -- node {$u$} (plant);
    \draw [->] (plant) -- node{$n_e$} (output);
    \draw [->] (min) -- node{$-$} (sum);
    \draw [->] (sum) -- node {$x$} (sat);
    \draw [->] (sat) -- node {sat$(x)$} (neuron);
    \draw [->] (plus) -- node {$+$} (sum);
    \draw [-] (tmp1) -- (tmp2) -| (min);

\end{tikzpicture}
  \caption{System with saturation block to limit the input to the controller. With reference input $r$, plasma density $n_e$, error $x$ and control action $u$.
}
\label{fig:block_diagram_with_saturation}
\end{figure}
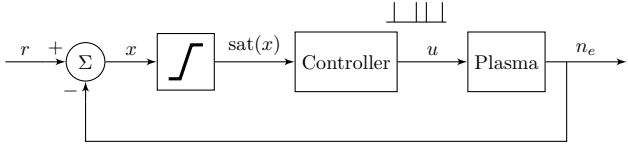

We explore adding an input saturation block, as shown in Fig. \ref{fig:block_diagram_with_saturation}, a solution also implemented in the AUG pellet controller \cite{Ploeckl-SDM}.
In the hybrid model, this changes the integrator dynamics of \eqref{eq:neuron_flow} to
\begin{equation}
    \dot{\xi} = \text{sat}(x) := 
    \begin{cases}
        0  & \text{if} \quad    x < 0 \\
        x & \text{if} \quad     0 \leqslant x \leqslant\frac{\Delta}{T_c}\\
        \frac{\Delta}{T_c}  & \text{if} \quad  \frac{\Delta}{T_c}<x.
    \end{cases}
    \label{eq:flow_with_saturation}
\end{equation}
This adjustment restricts the integrator's maximum increase to $\Delta$ per centrifuge actuator rotation.
The flow map then becomes
\begin{equation}
    F_{\text{IC}}(q) := 
    \begin{pmatrix}
        -\frac{1}{\tau}(x - r) \\
        \text{sat}(x) \\
        1
    \end{pmatrix},
    \label{eq:flowmap_with_saturation}
\end{equation}
and the hybrid system of the SDM with input clipping is then given by
\begin{equation}
    \mathcal{H}_{\text{IC}}=
    \begin{cases}
        \hfill  \dot{q} \; = \; F_{\text{IC}}(q), & q \in  \mathcal{C} \\
        q^+  \in \;  G_{\text{SDM}}(q),  & q \in  \mathcal{D} \text{, }
    \end{cases}
    \label{eq:hybrid_system_ic}
\end{equation}
where the flow map $F_{\text{IC}}$ is defined by \eqref{eq:flowmap_with_saturation},
the flow set $\mathcal{C}$ by \eqref{eq:flowSet},
the jump set $\mathcal{D}$ by \eqref{eq:jumpset}-\eqref{eq:jumpsetD2},
and the jump map $G_{\text{SDM}}(q)$ by \eqref{eq:jump_map_G1_sigmadelta}-\eqref{eq:jump_map_G2_sigmadelta}.
Similar results of Proposition~\ref{prop:hybrid-model-basic-properties} can be proven, mutatis mutandis, for this setup $\mathcal{H}_{\text{IC}}$.
Moreover, the result of Theorem~\ref{theorem:main} extends to $\mathcal{H}_{\text{IC}}$, as in the following proposition.

\begin{prop}
\label{prop:inputclipping}
Consider hybrid system $\mathcal{H}_{\textnormal{IC}}$ in \eqref{eq:hybrid_system_ic}, 
with $x(0,0) \leqslant r$ and $\xi(0,0)=T(0,0)=0$,
with a desired reference value $r \in \mathbb{R}_{>0}$, jump size $\alpha \in \mathbb{R}_{>0}$, and $r>\alpha$.
For all 
\begin{equation}
    T_c \in \left(0,\frac{\tau}{2} \ln{\frac{r}{r-\alpha}}\right),   
    \label{eq:Tc_condition_SDM}
\end{equation}
select 
\begin{equation}
\Delta \in \left(0,\left(r-(r-\alpha)e^{\frac{2T_c}{\tau}} \right)T_c \right],
\label{eq:Delta_condition_SDM}
\end{equation}
and define $\tau_d := \tau\ln{(\frac{r}{r-\alpha})} \in \R_{>0}$. 
Then, any solution $q$ of system \eqref{eq:hybrid_system_ic} satisfies 
for all $(t,j) \in \textnormal{dom }q$,
\begin{gather}
\begin{aligned}
 -\alpha < x(t,j) \leqslant \gamma^{\left(\frac{t}{\tau_d} -1\right)}  x(0,0) +  \alpha , &\\
   \text{if } \;  x(0,0) &>0 \\
   \min\left(r-e^{-\frac{t}{\tau}}\left(r-x(0,0)\right),-\alpha\right) < x(t,j) \leqslant  \alpha , &\\
   \text{if } \;  x(0,0) &\leqslant 0 ,
\end{aligned}
\label{eq:ultimate_bound_input_clipping}
\hspace*{16pt}
\raisetag{2.2\baselineskip} 
\end{gather}
with $\gamma := \frac{r-\alpha}{r} \in (0,1)$.
Also, 
$\lim\limits_{t+j \to \infty} \sup |x(t,j)| \leqslant \alpha$.
\end{prop}

\begin{figure}
    \begin{center}
        \includegraphics[width=0.45\textwidth]{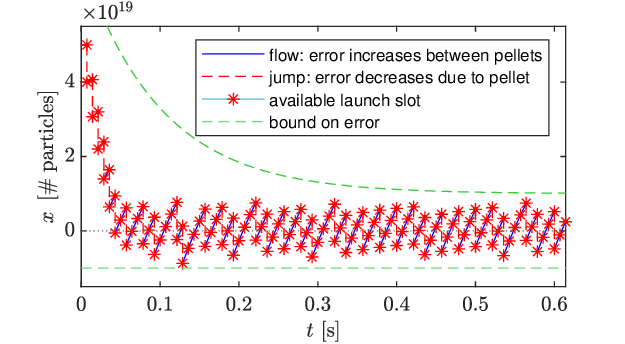} 
        \caption{Numerical simulation of the SDM controller with input clipping and a fast actuator.
        With $\alpha =1 \cdot 10^{19}$ particles/m$^3$, $r = 5 \cdot 10^{19}$~particles/m$^3$, $\tau = 0.1$~s, $T_c = 1/140$~s, $\Delta = 2.755\cdot10^{16}$ and $x(0,0)=r$. } 
        \label{fig:inputclipping1}
    \end{center}
\end{figure}

This approach has a practical stability property similar to the NM controller.
However, comparing \eqref{eq:Tc_condition} and \eqref{eq:Tc_condition_SDM}, we see that to guarantee the same ultimate bound $\alpha$, we need an actuator working at twice the speed.
Note that this is a theoretical guarantee, and it may be conservative in specific cases.
Fig.~\ref{fig:inputclipping1} confirms this in a simulation, using the values $\alpha =1 \cdot 10^{19}$ particles/m$^3$, $r = 5 \cdot 10^{19}$~particles/m$^3$, $\tau = 0.1$~s, as in Section~\ref{section:analysisNM}.
$T_c$ and $\Delta$ are within the constraints of Proposition~\ref{prop:inputclipping}, with $T_c = 1/140$~s (half of the time in Section \ref{section:analysisNM}), and $\Delta = 2.755\cdot10^{16}$ (upper limit).  
The system stays within the bounds of \eqref{eq:ultimate_bound_input_clipping}.

The proof of Proposition~\ref{prop:inputclipping} is given in Appendix~\ref{app:inputclipping}.
It is shown that due to the saturation in the dynamics of $\xi$ in \eqref{eq:flow_with_saturation}, when $x(t_i,i)\in [0,\frac{\Delta}{T_c})$\footnote{We use $t_i$ as the time at which jump $i \in \mathbb{Z}_{\geqslant 0}$ happens. At hybrid time $(t_i,i)$, $x(t_i,i)$ denotes the state after the jump. See Appendix~\ref{app:main} for more details.}
there exist cases where the error $x$ increases after a combined flow and pellet jump which influences the overall stability guarantees.
This leads to stricter conditions on $T_c$ and $\Delta$ in Proposition~\ref{prop:inputclipping}.
Where a small $\Delta$ for the NM controller in Theorem~\ref{theorem:main} leads to a pellet being launched every available slot, this is not a guarantee we have for the SDM in Proposition~\ref{prop:inputclipping}.
This means that for an $x(t_i,i) \in [0,\Delta/T_c)$ a scenario exists in which the controller requires at least 2 jumps (or rotations of the centrifuge) before a pellet is launched, and this is independent of the chosen $\Delta$, giving a more intuitive explanation for the doubling of the required actuator speed.

\begin{figure}
    \begin{center}
        \includegraphics[width=0.45\textwidth]{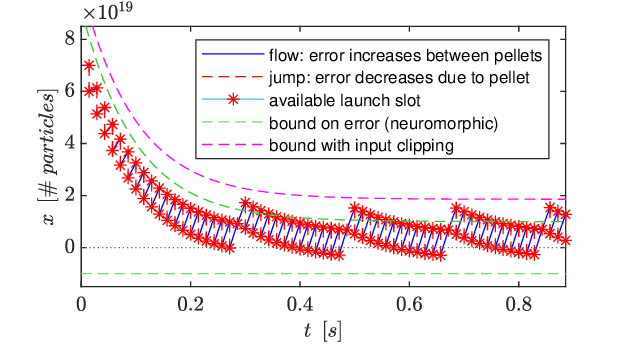} 
        \caption{Numerical simulation of SDM with input clipping and a slower actuator.
        With $\alpha =1 \cdot 10^{19}$ particles/m$^3$, $\tau = 0.1$~s, $T_c = 1/70$~s, $\Delta = 1$, $r=7 \times10^{19}$~particles/m$^3$ and $x(0,0)=r$.} 
        \label{fig:inputclipping}
    \end{center}
\end{figure}

It is of interest to compare the SDM with the same actuator limits as the NM controller, i.e., $T_c = \tau \ln{\frac{r}{r-\alpha}}$, the maximum allowed value in Theorem~\ref{theorem:main}.
In Remark~\ref{rem:SDM_same_speed_NM} in Appendix~\ref{app:inputclipping} it is shown that for $\Delta =1$ we can guarantee an ultimate bound of $\left(-\alpha,\alpha \left( 2-\frac{\alpha}{r}\right)\right]$ for the SDM with input clipping.
Fig. \ref{fig:inputclipping} illustrates this for a high reference value, using the values $\alpha =1 \cdot 10^{19}$ particles/m$^3$, $\tau = 0.1$~s, $T_c = 1/70$~s and $\Delta = 1$ as in Section \ref{section:analysisNM},
and with a higher reference value $r=7 \times10^{19}$~particles/m$^3$.
In steady-state, the system stays within the ultimate bound $\left(-\alpha,\alpha \left( 2-\frac{\alpha}{r}\right)\right]$ (magenta dashed line), which is wider than the bound $\left(-\alpha,\alpha \right]$ of the NM controller (green dashed line).
Comparing Fig.~\ref{fig:delta_min_max} and Fig.~\ref{fig:inputclipping}, we see no difference in transient behaviour, where all available timeslots are used.
However, the simulations confirm the need for a faster actuator to stay within the ultimate bound $\alpha$ in the steady-state phase.
In Section~\ref{subsec:comparison} we compare the NM controller and SDM in more detail, but first we present an alternative solution to the wind-up problem discussed in Section~\ref{subsection:hybridmodel_SDM} for SDM.

\subsection{Adjusting the jump map}
\label{subsec:SDM_jm}
An alternative solution (without input clipping) to avoid the integrator wind-up, is a reset condition that subtracts multiples of $\Delta$ from the neuron membrane potential $\xi$ until it is below $\Delta$ again.
This method keeps the benefit of having an error memory in steady-state operation, without having the undershoot problems.
Instead of the standard reset for SDM in \eqref{eq:reset_neuron_SDM}, we use
\begin{equation}
    \xi^+ = \xi - k\Delta,
    \label{eq:reset_neuron_SDM2}
\end{equation}
with $k= \Bigl \lfloor \frac{\xi}{\Delta} \Bigr \rfloor$, 
the largest integer smaller than or equal to $\frac{\xi}{\Delta}$. 
The jump map then becomes
\begin{equation}
    G_{\text{JM}}(q) := G_{\text{JM}1}(q) \cup G_{\text{JM2}}(q),
    \label{eq:jump_map_sigmadelta2}
\end{equation}
with
\begin{equation}
    G_{\text{JM}1}(q) := 
    \begin{cases}
    \left\{
        \begin{pmatrix}
            x \\
            \xi \\
            0
        \end{pmatrix} 
        \right\}
        & q \in \mathcal{D}_1 \\
        \hfill \emptyset \hfill  & q \notin \mathcal{D}_1
    \end{cases}
    \label{eq:jump_map_G1_sigmadelta2}
\end{equation}
\begin{equation}
    G_{\text{JM}2}(q) := 
    \begin{cases}
    \left\{
        \begin{pmatrix}
            x - \alpha \\
            \xi - k\Delta \\
            0
        \end{pmatrix} 
        \right\}
        & q \in \mathcal{D}_2 \\
        \hfill \emptyset \hfill  & q \notin \mathcal{D}_2,
    \end{cases}
    \label{eq:jump_map_G2_sigmadelta2}
\end{equation}
where $\mathcal{D}_1$ and $\mathcal{D}_2$ are defined in the same way as for the NM controller in \eqref{eq:jumpsetD1}-\eqref{eq:jumpsetD2}.

We now define the hybrid system of the SDM with the adjusted jump map as
\begin{equation}
    \mathcal{H}_{\text{JM}}=
    \begin{cases}
        \hfill  \dot{q} \; = \; F(q), & q \in  \mathcal{C} \\
        q^+  \in \;  G_{\text{JM}}(q),  & q \in  \mathcal{D} \text{, }
    \end{cases}
    \label{eq:hybrid_system_k}
\end{equation}
where the flow map $F$ is defined by \eqref{eq:flowmap_with_saturation},
the flow set $\mathcal{C}$ by \eqref{eq:flowSet},
the jump set $\mathcal{D}$ by \eqref{eq:jumpset}-\eqref{eq:jumpsetD2},
and the jump map $G_{\text{JM}}$ by \eqref{eq:jump_map_G1_sigmadelta2}-\eqref{eq:jump_map_G2_sigmadelta2}.
Similar results of Proposition~\ref{prop:hybrid-model-basic-properties} can be proven, mutatis mutandis, for system $\mathcal{H}_{\text{JM}}$.
Moreover, the result of Theorem~\ref{theorem:main} extends to $\mathcal{H}_{\text{JM}}$, as in the following proposition.

\begin{prop}
\label{prop:jumpmap}
Consider hybrid system $\mathcal{H}_{\textnormal{JM}}$ in \eqref{eq:hybrid_system_k}, 
with $x(0,0) \leqslant r$ and $\xi(0,0)=T(0,0)=0$,
with a desired reference value $r \in \mathbb{R}_{>0}$, jump size $\alpha \in \mathbb{R}_{>0}$, and $r>\alpha$.
For all
\begin{equation}
    T_c \in \left(0,\tau \ln{\frac{r}{r-\alpha}}\right],
    \label{eq:Tc_condition_JM}
\end{equation}
select 
\begin{align}
    \Delta \in \left(0,r \tau \ln{\frac{r}{r-\alpha}} - r \tau \left(1- \frac{r-\alpha}{r} e^{\frac{T_c}{\tau}}\right) - r T_c\right]
    \label{eq:Delta_condition_JM}
    \hspace*{25pt}
    \raisetag{3\baselineskip} 
\end{align}
and define $\tau_d := \tau\ln{(\frac{r}{r-\alpha})} \in \R_{>0}$. 
Then, any solution $q$ of the system \eqref{eq:hybrid_system_k} with $\xi(0,0)=0$, satisfies for all $(t,j) \in \textnormal{dom }q$, 
\begin{gather}
\begin{aligned}
 -\alpha < x(t,j) \leqslant \gamma^{\left(\frac{t}{\tau_d} -1\right)}  x(0,0) + \alpha , &\\
   \text{if } \;  x(0,0) &>0 \\
   \min\left(r-e^{-\frac{t}{\tau}}\left(r-x(0,0)\right),-\alpha\right) < x(t,j) \leqslant  \alpha , &\\
   \text{if } \;  x(0,0) &\leqslant 0 ,
\end{aligned}
\label{eq:ultimate_bound_JM}
\hspace*{16pt}
\raisetag{2.2\baselineskip} 
\end{gather}
with $\gamma := \frac{r-\alpha}{r} \in (0,1)$.
Also, 
$\lim\limits_{t+j \to \infty} \sup |x(t,j)| \leqslant \alpha$.
\end{prop}

For this system we obtain the same stability properties as for the NM controller in \eqref{eq:ultimate_bound}, 
as \eqref{eq:Tc_condition_JM}-\eqref{eq:ultimate_bound_JM} are equivalent to \eqref{eq:Tc_condition}-\eqref{eq:ultimate_bound} for the same parameters.
This is shown in Appendix \ref{app:sdm2}.
In the next section we compare the NM controller to SDM in more detail.

\begin{figure}
    \begin{center}
        \includegraphics[width=0.47\textwidth]{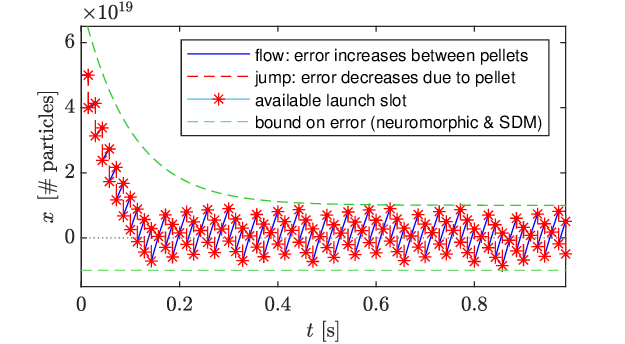} 
        \caption{Numerical simulation of SDM with adjusted jump map.
        Using the same values $\alpha =1 \cdot 10^{19}$ particles/m$^3$, $r=5 \times10^{19}$~particles/m$^3$, $\tau = 0.1$~s, $T_c = 1/70$~s, $\Delta = 1.569 \cdot 10^{16}$ and $x(0,0)=r$ as in Fig. \ref{fig:delta_min_max}.
        The system reaches a limit cycle with the same bounds as the neuromorphic controller.} 
        \label{fig:sdm2}
    \end{center}
\end{figure}

\subsection{Comparing SDM to NM control}
\label{subsec:comparison}

To obtain favourable transient and steady-state behaviour, we compare the NM controller with the SDM using the adjusted jump map from Section~\ref{subsec:SDM_jm}. 
The main theorem and derived propositions show that, for the same actuator, NM and SDM have the same guaranteed ultimate bound.
Although SDM was expected to achieve a smaller bound due to timing corrections via the residue in 
$\xi$, comparison of Proposition~\ref{prop:jumpmap} with Theorem~\ref{theorem:main} shows that this provides no advantage over NM, which resets $\xi$ to zero. 
This is confirmed by the simulations in Fig.~\ref{fig:delta_min_max} and \ref{fig:sdm2}, where NM has a tighter bound for identical parameters. 
This can be understood by viewing the NM reset as a particular realization of SDM: the sequence 
$\xi^+=\xi-\Delta=0$ represents the worst timing case, producing the longest interval between pellet injections and coinciding with nominal NM behaviour. 
Since stability analysis must account for this case, SDM offers no intrinsic theoretical advantage. 
Moreover, SDM requires mitigation of integrator wind-up, whereas NM does not. 
Therefore, the proposed NM controller is simpler while providing equal guarantees.

\section{Pellet preparation time}
\label{section:refr}
For both the centrifuge and the gas gun type actuator, there is a pellet preparation time.
This can be implemented similar to a refractory period in the neuron model \cite{Gerstner}, as the controller cannot generate a new spike until this waiting time is over.
We consider both the NM controller and SDM as described in Section~\ref{subsec:SDM_jm} and
implement the pellet preparation time by adding an extra timer
$    \dot{T}_p = 1,
    \label{eq:timer_refractory}$
where $T_p \in \mathbb{R}_{\geqslant 0}$.
When a pellet is fired, the timer is reset to zero, with discrete jump dynamics
$    T_p^+ = 0 .
    \label{eq:reset_timer_refr}$
This leads to an extra variable in the state $q$ of the hybrid system in \eqref{eq:hybrid_system_general}, now defined as $\tilde q := (x,\xi,T,T_p) \in \mathcal{\tilde X}$, with
$\mathcal{\tilde X}:=\{(x,\xi,T,T_p) \in \mathbb{R}^4 : (x \leqslant r) \wedge (\xi \geqslant 0) \wedge (T \geqslant 0) \wedge (T_p \geqslant 0) \}$.
The flow map $F_{\text{PP}}$ becomes 
\begin{equation}
    F_{\text{PP}}(\tilde q) := 
    \begin{pmatrix}
        -\frac{1}{\tau}(x -r) \\
        \max(0,x) \\
        1 \\
        1
    \end{pmatrix} \quad q \in \mathcal{C}_{\text{PP}},
    \label{eq:flowmap_PP}
\end{equation}
where the flow set $\mathcal{C}_{\text{PP}}$ is given by
\begin{equation}
    \mathcal{C}_{\text{PP}} := \{ \tilde q \in \mathcal{\tilde X}: T \leqslant T_c \}.
    \label{eq:flowSetPP}
\end{equation}
reset, where we check if
\begin{equation}
    T_p \geqslant T_{\text{prep}} ,
    \label{eq:jump_cond_timer_PP}
\end{equation}
with $T_{\text{prep}} \in \R_{\geqslant 0}$ the pellet preparation time.
The jump set is defined as
\begin{equation}
    \mathcal{D}_{\text{PP}} := \{ \tilde q \in \mathcal{\tilde X} : T \geqslant T_c \}, \label{eq:jumpset_PP}
\end{equation}
which we rewrite as $\mathcal{D}_{\text{PP}} := \mathcal{D}_{\text{PP}1} \cup \mathcal{D}_{\text{PP}2}$, with
\begin{equation}
\begin{aligned}
    \mathcal{D}_{\text{PP}1} &:= \{ \tilde q \in \mathcal{\tilde X}: (T \geqslant T_c ) \wedge ( (\xi \leqslant \Delta ) \vee (T_p\leqslant T_{\text{prep}}) )\} \\
    \mathcal{D}_{\text{PP}2} &:= \{\tilde q \in \mathcal{\tilde X}: (T \geqslant T_c ) \wedge ( \xi \geqslant \Delta ) \wedge(T_p \geqslant T_{\text{prep}}) \}.
    \label{eq:jumpsetD2_PP}
\end{aligned}
\end{equation}
The jump map $G_{\text{PP}}$ becomes
\begin{equation}
    G_{\text{PP}}(\tilde q) \in G_{\text{PP}1}(\tilde q) \cup G_{\text{PP}2}(\tilde q),
    \label{eq:jump_map_PP}
\end{equation}
with
\begin{equation}
    G_{\text{PP}1}(\tilde q) := 
    \begin{cases}
    \left\{
        \begin{pmatrix}
            x \\
            \xi \\
            0 \\
            T_p
        \end{pmatrix} 
        \right\}
        & \tilde q \in \mathcal{D}_{\text{PP}1} \\
        \hfill \emptyset \hfill  & \tilde q \notin \mathcal{D}_{\text{PP}1}
    \end{cases}
    \label{eq:jump_map_G1_PP}
\end{equation}
\begin{equation}
    G_{\text{PP}2}(\tilde q) \in 
    \begin{cases}
    \left\{
        \begin{pmatrix}
            x - \alpha \\
            \xi - \varepsilon \\
            0 \\
            0
        \end{pmatrix} 
        \right\}
        & \tilde q \in \mathcal{D}_{\text{PP}2} \\
        \hfill \emptyset \hfill  & \tilde q \notin \mathcal{D}_{\text{PP}2} ,
    \end{cases}
    \label{eq:jump_map_G2_PP}
\end{equation}
with $\varepsilon \in \{\xi, k\Delta \}$ and $k= \Bigl \lfloor \frac{\xi}{\Delta} \Bigr \rfloor$.
Clearly, $\varepsilon = \xi$ leads to the NM controller presented in Section~\ref{section:NM}-\ref{section:analysisNM}, and $\varepsilon = k\Delta$ to the SDM with adjusted jump map in Section~\ref{subsec:SDM_jm}.
We define the hybrid system of the controller with a pellet preparation time as
\begin{equation}
    \mathcal{H}_{\text{PP}}=
    \begin{cases}
        \hfill  \dot{\tilde q} \; = \; F_{\text{PP}}(\tilde q), & \tilde q \in  \mathcal{C}_{\text{PP}} \\
        \tilde q^+  \in \;  G_{\text{PP}}(\tilde q),  & \tilde q \in  \mathcal{D}_{\text{PP}} \text{, }
    \end{cases}
    \label{eq:hybrid_system_PP}
\end{equation}
with $F_{\text{PP}},\mathcal{C}_{\text{PP}},G_{\text{PP}},\mathcal{D}_{\text{PP}}$ as in \eqref{eq:flowmap_PP}-\eqref{eq:jump_map_G2_PP}.
Similar results as in Proposition~\ref{prop:hybrid-model-basic-properties} can be proven, mutatis mutandis, for system $\mathcal{H}_{\textnormal{PP}}$.
Moreover, the result of Theorem~\ref{theorem:main} extends to $\mathcal{H}_{\text{PP}}$, as in the following proposition.

\begin{prop}
\label{prop:refractory}
Consider hybrid system $\mathcal{H}_{\textnormal{PP}}$ in \eqref{eq:hybrid_system_PP}, 
with $x(0,0) \leqslant r$ and $\xi(0,0)=T(0,0)=T_p(0,0)=0$,
with a desired reference value $r \in \mathbb{R}_{>0}$, jump size $\alpha \in \mathbb{R}_{>0}$, with $r>\alpha$, 
and pellet preparation time $T_{\textnormal{prep}} \in \R_{\geqslant0}$ with 
$l= \Bigl \lceil \frac{T_{\textnormal{prep}}}{T_c} \Bigr \rceil \in \mathbb{N}$.
If 
\begin{equation}
    T_{\textnormal{prep}} \in \left(0, \tau \ln{\frac{r}{r-\alpha}}\right],    
    \label{eq:Tprep_condition_PP}
\end{equation}
then for all 
\begin{equation}
    T_c \in \left(0, \frac{\tau}{l} \ln{\frac{r}{r-\alpha}}\right],    
    \label{eq:Tc_condition_PP}
\end{equation}
select 
\begin{align}
\Delta &\in \left(0,r \tau \ln{\frac{r}{r-\alpha}} - r \tau \left(1- \frac{r-\alpha}{r} e^{\frac{l T_c}{\tau}}\right) - r l T_c\right]
\label{eq:Delta_condition_PP}
\hspace*{19pt}
\raisetag{3\baselineskip} 
\end{align}
and define $\tau_d := \tau\ln{(\frac{r}{r-\alpha})} \in \R_{>0}$. 
Then, any solution $\tilde q$ of the system \eqref{eq:hybrid_system_PP} satisfies for all $(t,j) \in \textnormal{dom }\tilde q$, 
\begin{gather}
\begin{aligned}
 -\alpha < x(t,j) \leqslant \gamma^{\left(\frac{t}{\tau_d} -1\right)}  x(0,0) + \alpha , &\\
   \text{if } \;  x(0,0) &>0 \\
   \min\left(r-e^{-\frac{t}{\tau}}\left(r-x(0,0)\right),-\alpha\right) < x(t,j) \leqslant  \alpha , &\\
   \text{if } \;  x(0,0) &\leqslant 0 ,
\end{aligned}
\label{eq:ultimate_bound_PP}
\hspace*{16pt}
\raisetag{2.2\baselineskip} 
\end{gather}
with $\gamma := \frac{r-\alpha}{r} \in (0,1)$.
Also, 
$\lim\limits_{t+j \to \infty} \sup |x(t,j)| \leqslant \alpha$.
\end{prop}

The proof of Proposition~\ref{prop:refractory} is given in Appendix~\ref{app:refr}.
When $T_{\textnormal{prep}} \leqslant T_c$, we get $l=1$ and the pellet preparation time does not introduce any new behaviour in the system,
hence the results in Section~\ref{section:analysisNM} and \ref{subsec:SDM_jm} are valid.
When $T_{\textnormal{prep}} \geqslant T_c$, the minimum time between two pellets is determined as $lT_c$, with $l= \Bigl \lceil \frac{T_{\textnormal{prep}}}{T_c} \Bigr \rceil \in \mathbb{N}$, the smallest integer larger than or equal to $\frac{T_{\textnormal{prep}}}{T_c}$.
Similar to \eqref{eq:r_max}, we can establish a maximum reference
\begin{equation}
    r_\textnormal{max} = \frac{e^{\frac{lT_c}{\tau}}}{e^{\frac{lT_c}{\tau}}-1} \alpha \in \R_{\geqslant 0}.
    \label{eq:r_max_PP}
\end{equation}
As long as we stay below this limit, the system will converge to the ultimate bound of Proposition~\ref{prop:refractory}.

\begin{rem}
\label{rem:gas_gun}
    We can extend Proposition~\ref{prop:refractory} to a practical stability property for a gas gun type actuator by considering $T_\textnormal{prep}$ satisfies \eqref{eq:Tprep_condition_PP}, $T_c$ now being the sample time of the actuator, which is usually very small compared to $T_\text{prep}$, and $lT_c=T_\text{prep}$ in \eqref{eq:Delta_condition_PP}.
    The ultimate bound in \eqref{eq:ultimate_bound_PP} can further be refined, depending on the value of $r$.
    Determining the exact ultimate bound is left for future work.
\end{rem}

\begin{rem}
\label{rem:AUG}
The SDM control schemes as presented here are 
capturing the main characteristics of the AUG controller \cite{Ploeckl-SDM} in our SDM setup, although they are not exactly the same.
The SDM variant of \cite{Ploeckl-SDM} considered here, show that SDM can lead to a similar ultimate bound as NM control, given certain constraints. 
Although the stability properties are similar, SDM is more complicated to analyse and implement because of the wind-up mitigation, which is also present in the SDM control scheme at AUG.
In addition, the control scheme at AUG has proportional and integral gain in the feedback loop \cite{Kudlacek-overview}.
While the effect of the proportional gain can be linked directly to the tuning of the threshold $\Delta$, analysing the effect of the integrator is more difficult.
Adding an integrator can remove the average steady-state error we see in, e.g., Fig.~\ref{fig:delta_min_max}, but it is not guaranteed to narrow the ultimate bound.
Given the above analysis, if an average steady-state error within the ultimate bound is acceptable, there is no benefit in adding an integrator.
Even more, improper tuning of an integrator, can lead to overshoot and undesirable transient behaviour.
\end{rem}

\section{Conclusion}
\label{section:conclusion}
This work addresses the problem of regulating plasma density in tokamaks using discrete pellet injection, with a focus on ensuring practical closed-loop stability under actuator limitations, including synchronisation to the centrifuge and a pellet preparation time.
The newly proposed NM controller achieves the same practical stability guarantees as SDM approaches, but without suffering from integrator wind-up. 
As a result, SDM-specific mitigation strategies, such as input clipping or reset modification, are not required to maintain performance.
In practical terms, this means that we can adopt the NM controller to obtain a solution that is simpler to implement and easier to analyse, without sacrificing stability guarantees. 
Since SDM-based designs require additional modifications and design constraints without improving practical stability, they do not provide a clear benefit for this application.

The analysis in this paper was carried out using a simplified plasma density model and primarily considered actuator dynamics through pellet preparation time. 
Further work is required to assess robustness under more realistic operating conditions, including time-varying reference trajectories, communication delays and pellet losses, and deployment in high-fidelity simulation environments or experimental tokamak settings.

\appendix
\input{appendix}

\begin{ack}            
This publication is part of the project Balls-to-the-wall: Keeping Hydrogen Burning with Ice Bullets (Project No. 19695) of the NWO Talent Programme VIDI, financed partly by the Dutch Research Council (NWO).
DIFFER is part of the institutes organization of NWO.\\
This work has been carried out within the framework of the EUROfusion Consortium, funded by the European Union via the Euratom Research and Training Programme (Grant Agreement No 101052200 - EUROfusion). Views and opinions expressed are however those of the author(s) only and do not necessarily reflect those of the European Union or the European Commission. Neither the European Union not the European Commission can be held responsible for them.
\end{ack}

\end{document}

%% file: appendix.tex
\section{Proof of Proposition~\ref{prop:hybrid-model-basic-properties}}
\label{app:basic_properties}
Consider system \eqref{eq:hybrid_system_general} with \eqref{eq:flowmap}-\eqref{eq:jump_map_G2}. We prove the three items separately. 

\noindent\emph{(i)} Flow and jump sets $\mathcal{C}$ and $\mathcal{D}$ in \eqref{eq:flowSet}-\eqref{eq:jumpsetD2} are closed subsets of $\mathcal{X}$. In addition, $F$ and $G$ are continuous.
Hence the system satisfies the hybrid basic conditions.

\noindent \emph{(ii)} Using the particular structure of the hybrid model, the timer $T$ imposes a minimum dwell-time $T_c$, i.e., for all $(s,i),(t,j) \in \textnormal{dom } q$ with $s+i \leqslant t+j$ we have $j-i \leqslant \frac{t-s}{T_c}+1$.
This precludes Zeno behaviour, i.e., infinite amount of jumps in a finite amount of time.

\noindent \emph{(iii)} 
We verify the desired result by ensuring that the conditions given in \cite[Proposition~6.10]{Goebel:hybrid} hold. 
Let $q = (x, \xi, T) \in \mathcal{C}$.
If $q$ is in the interior of $\mathcal{C}$, i.e., $T<T_c$,
the tangent cone to $\mathcal{C}$ is given by $T_\mathcal{C}(q) = \R \times \R_{\geqslant 0} \times \R_{\geqslant 0}$. 
If $q$ is on the boundary of $\mathcal{C}$, i.e., 
$T=T_c, T_\mathcal{C}(q)=\R \times \R_{\geqslant 0} \times \{0\}$
In view of the definition of $F$ in \eqref{eq:flowmap} and of $\mathcal{D}$ in \eqref{eq:jumpset}, we observe that, for any $q \in \mathcal{C}\setminus\mathcal{D}, F(q) \in T_\mathcal{C}(q)$. 
Moreover, it can also be seen that for each $q \in \mathcal{C}\setminus\mathcal{D}$, there exists a neighbourhood $\mathcal{U}$ of $q$ such that $\mathcal{U} \cap \mathcal{C} \cap \mathcal{D} = \emptyset$. 
Given the last two observations, we deduce that \cite[Proposition~6.10~(VC)]{Goebel:hybrid} holds. 
On the other hand, we have that \cite[Proposition~6.10(b)]{Goebel:hybrid} cannot occur as $F$ is linear. 
Finally, $G(\mathcal{D}) \subset \mathcal{C}$ so that  \cite[Proposition~6.10(c)]{Goebel:hybrid} cannot occur.
We conclude that maximal solutions are complete by application of \cite[Proposition~6.10]{Goebel:hybrid}.

Combining this with \emph{(ii)}, we can conclude Proposition~\ref{prop:hybrid-model-basic-properties}\emph{(iii)} holds, maximal solutions are $t$-complete, i.e., $\sup_t  \textnormal{dom } q = +\infty$.
\hfill $\blacksquare$

\section{Proof of Theorem~\ref{theorem:main}}
\label{app:main}
Consider system \eqref{eq:hybrid_system_general} with \eqref{eq:flowmap}-\eqref{eq:jump_map_G2} and let $q$ be a maximal solution with $x(0,0) \leqslant r$.
In view of Proposition~\ref{prop:hybrid-model-basic-properties}, $q$ is $t$-complete.
Pick any $(t,j) \in \text{dom } q$ and let 
$0 = t_0 \leqslant t_1 \leqslant \ldots \leqslant t_{j+1} = t$
satisfy $\text{dom }q \ \cap \ ([0,t] \times \{0,1,\ldots,j\}) = \bigcup_{i=0}^j [t_i,t_{i+1}] \times \{i\}$.
Note that $t_{j+1}=t$ is not necessarily a jump time but $t_1,\ldots,t_j$ are jump times.
To separate the jumps where a pellet is being fired, from the jumps where no pellet is fired, 
we define the set of hybrid times at which a jump occurs 
due to a spike generated by jump set $\mathcal{D}_2$, i.e., due to a pellet being fired, as
$    \mathcal{P}(q) := \{(t, i) \in \text{dom } q : q(t,i) \in \mathcal{D}_2 \text{ and } \\
q(t, i + 1) \in G_2(q(t, i)) \}$.

\noindent \textbf{General expressions of solutions}\\
We now compute the expression of the solution for the state and for the neuron membrane potential.
For all $i \in \{0,1,\ldots,j\}$ and all $s \in [t_i,t_{i+1}]$,
$q(s,i) \in \mathcal{C}$  
and from \eqref{eq:openloop_x}, we have 
\begin{equation}
    x(s,i) = r-e^{-\frac{1}{\tau}(s-t_i)}\left(r-x(t_i,i)\right).
    \label{eq:x_s_i}
\end{equation} 
Similarly, for each 
$i \in \{0,1,\ldots,j\}$ and all $s \in [t_i,t_{i+1}]$,
from the neuron dynamics \eqref{eq:neuron_flow}, we have
\begin{equation}
    \xi(s,i) = \xi(t_i,i) + \int_{t_i}^{s} \text{max}\left(0,x(\tilde{s},i)\right)d\tilde{s}.
    \label{eq:xi_s_i}
\end{equation}
First we consider the case where $x(t_i,i) > 0$.
In this case, from \eqref{eq:xi_s_i}, we have
for all $s \in [t_i,t_i+1]$,
\begin{equation}
    \xi(s,i) = \xi(t_i,i) + \int_{t_i}^{s} x(\tilde{s},i) d\tilde{s} .
    \label{eq:xi_s_i_zonder_max}
\end{equation}
The case $x(t_i,i) \leqslant 0$ will be considered later and in the next part of the proof we focus on the case $x(t_i,i) >0$.

On the other hand, using the jump map in \eqref{eq:jump_map}-\eqref{eq:jump_map_G2}, we have that for each $k \in \mathbb{Z}_{> 0}$, with $(t_k, k-1) \in \text{dom }q$ and $(t_k, k) \in \text{dom } q$,
\begin{gather}
    \begin{aligned}
        \xi(t_k,k) &= \xi(t_k,k-1) &\text{ when } (t_k,k-1) \notin \mathcal{P}(q) \hspace*{29pt}\\
        \hfill \xi(t_k,k) &= 0 \hfill &\text{ when } (t_k,k-1) \in \mathcal{P}(q).
        \hspace*{29pt}
        \raisetag{1.4\baselineskip} 
    \end{aligned}
    \label{eq:xi_jumps}
\end{gather}

\noindent \textbf{Increase of $x$ between two pellet launches}\\
Using the expressions we compute, our goal is now to determine the increase of $x$ between two pellet launches.
Using jump set $\mathcal{D}_2$ from \eqref{eq:jumpsetD2}, define
\begin{align}
    p(i) := \text{inf}\{ \tilde{i} \geqslant i : \exists s \in \mathbb{R}_{\geqslant 0}
    \text{ such that } (s,\tilde{i}) \in \mathcal{P}(q) \}.
    \label{eq:pellet_jump_time}
\end{align}
Note that $t_{p(i)}$ is the first time after $t_i$ at which a pellet is fired,
and that, between $t_i$ and $t_{p(i)}$ only jumps generated by the timer can occur, where no pellet is fired.
To consider the increase of $x$ between 2 pellet launches, we start from a jump with a pellet launch.
For each $i \in \{1,2,\ldots,j\}$, let $(t_i,i-1) \in \mathcal{P}(q)$ and consider that there are $n \in \mathbb{Z}_{\geqslant 0}$ jumps without a pellet between $t_i$ and $t_{p(i)}$.
The case where $i=0$ will be considered later.
From \eqref{eq:xi_s_i_zonder_max} we have that for all $m \in \{0,1,\ldots, n-1\}$ and all $s \in [t_{i+m},t_{i+m+1}]$,
\begin{equation}
    \xi(s,i+m) = \xi(t_{i+m},i+m) + \int_{t_{i+m}}^{s} x(\tilde{s},i+m) d\tilde{s}.
    \label{eq:xi_flow_tussenin}
\end{equation}
Note that in case $n=0$, we get $m=0$.
Similarly, for all $s \in [t_{i+n},t_{p(i)}]$,
\begin{equation}
    \xi(s,i+n) = \xi(t_{i+n},i+n) + \int_{t_{i+n}}^{s} x(\tilde{s},i+n) d\tilde{s}.
    \label{eq:xi_flow_voor_pellet_spike}
\end{equation}
Moreover, recalling from \eqref{eq:xi_jumps} that when the jump time $(t_k,k-1) \notin \mathcal{P}(q)$, $\xi(t_k,k) = \xi(t_k,k-1)$, from \eqref{eq:xi_flow_tussenin} and \eqref{eq:xi_flow_voor_pellet_spike} we obtain, for all $s \in [t_i,t_{p(i)}]$,
    $\xi(s,l) = \xi(t_{i},i) + \int_{t_i}^{s} x(\tilde{s},l) d\tilde{s}$.
with $(s,l) \in \text{dom } q$ for some $l \in \{i,i+1,\ldots,i+n\}$.
In the sequel, we will sometimes write $\xi(s,\cdot)$ for $\xi(s,l)$, in such cases, when $l$ is clear from the context.
Moreover, since for $(t_i,i-1) \in \mathcal{P}(q)$, from \eqref{eq:xi_jumps}, we have 
$\xi(t_i,i) = 0$ and thus, for all $s \in [t_i,t_{p(i)}]$,
\begin{equation}
    \xi(s,\cdot) = \int_{t_i}^{s} x(\tilde{s},\cdot)d\tilde{s} .
    \label{eq:xi_total_flow}
\end{equation}
Using the jump map in \eqref{eq:jump_map}-\eqref{eq:jump_map_G2}, we have that for each $k \in \mathbb{Z}_{> 0}$, with $(t_k, k-1) \in \text{dom }q$ and $(t_k, k) \in \text{dom } q$, 
\begin{gather}
    \begin{aligned}
        x(t_k,k) &= x(t_k,k-1) \; \text{ when } (t_k,k-1) \notin \mathcal{P}(q) \\
        \hfill x(t_k,k) &= x(t_k,k-1) - \alpha \; \text{ when } (t_k,k-1) \in \mathcal{P}(q)
    \end{aligned}
    \label{eq:reset_x_T2}
    \hspace*{19pt}
    \raisetag{1.9\baselineskip} 
\end{gather}
We have from \eqref{eq:x_s_i} that for all $m \in \{0,1,\ldots,n-1\}$ and all $s \in [t_{i+m},t_{i+m+1}]$,
\begin{align}
    x(s,i+m) = r-e^{-\frac{1}{\tau}(s-t_{i+m})}\left(r-x(t_{i+m},i+m)\right).
    \label{eq:x_between_jumps}
\end{align} 
For all $s \in [t_{i+n},t_{p(i)}]$,
\begin{align}
    x(s,i+n) = r-e^{-\frac{1}{\tau}(s-t_{i+n})}\left(r-x(t_{i+n},i+n)\right).
    \label{eq:x_between_last_jumps}
    \hspace*{27pt}
    \raisetag{0.8\baselineskip} 
\end{align}
So, from \eqref{eq:reset_x_T2}-\eqref{eq:x_between_last_jumps}, we obtain for all $s \in [t_i,t_{p(i)}]$,
\begin{equation}
    x(s,\cdot) = r-e^{-\frac{1}{\tau}(s-t_i)}\left(r-x(t_i,i)\right).
    \label{eq:x_flow_between_pellets}
\end{equation}
From \eqref{eq:pellet_jump_time} and \eqref{eq:x_flow_between_pellets}, we get
\begin{equation}
    x(t_{p(i)},p(i)-1) = r-e^{-\frac{1}{\tau}(t_{p(i)}-t_i)}\left(r-x(t_i,i)\right),
    \label{eq:V_increase_between_pellets}
\end{equation}
which is the increase in $x$ between two pellet jumps (note that timer jumps without a pellet do not cause any discontinuities in $x$), without the effect of the last pellet injection.

Combining \eqref{eq:xi_total_flow} with \eqref{eq:x_flow_between_pellets}, we get for all $s \in [t_i,t_{p(i)}]$,
\begin{gather}
\begin{aligned}
    \xi(s,\cdot) &= \int_{t_i}^{s} \left(e^{-\frac{1}{\tau}(\tilde{s}-t_i)}\left(x(t_i,i)-r\right)+r\right)d\tilde{s} \\
     &= \tau (x(t_i,i)-r) \left(1-e^{-\frac{1}{\tau}(s-t_i)}\right)+r(s-t_i).
     \label{eq:neuron_between_pellets}
\end{aligned}
    \hspace*{21pt}
    \raisetag{2.4\baselineskip} 
\end{gather}
Note that $i=0$ is not considered above since there is no pellet launch at $t_0$, given the initial conditions.
However, recalling that $t_0=0$ and from $T(0,0)=\xi(0,0)=0$, \eqref{eq:neuron_between_pellets} holds for $i\in \{0,1,\ldots,j\}$.

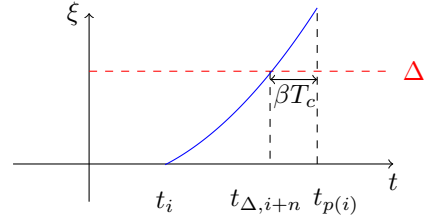
\begin{figure}[thpb]
      \centering
    \begin{tikzpicture}
      \draw[->] (-1, 0) -- (4, 0) node[below] {$t$};
      \draw[->] (0, -0.5) -- (0, 2) node[left] {$\xi$};
      \draw[<->] (2.38,1.12) -- (3, 1.12) node[above] {};
      \draw[dashed,red] (0, 1.23) -- (4, 1.23) node[right] {$\Delta$};
      \draw[dashed] (2.38, 1.23) -- (2.38, 0) node[right] {};
      \draw[dashed] (3, 2) -- (3, 0) node[right] {};
      \node (tmp1) at (1,-0.5) {$t_i$};
      \node (tmp2) at (2.33,-0.5) {$t_{\Delta, i+n}$};
      \node (tmp3) at (3.25,-0.5) {$t_{p(i)}$};
      \node (tmp4) at (2.71,0.9) {$\beta T_c$};
      \draw[domain=1:3, smooth, variable=\x, blue] plot ({\x}, {0.26*\x*\x-0.27});
    \end{tikzpicture}
     \caption{At time $t_i$, $\xi=0$, when $\xi$ reaches the threshold $\Delta$ at time $t_{\Delta, i+n}$, it needs to wait for the timer $T$ to reach $T_c$ before triggering a jump at $t_{p(i)}$.}
      \label{fig:xi_overshoot}
   \end{figure}

\noindent \textbf{Computing maximum time between two pellets launches}\\
To capture the effect of synchronising the neuron spike to the timer, we define
$t_{\Delta, i+n} := \inf\{s\geqslant t_{i+n} : \xi(s, i+n) = \Delta\} $
as the time the neuron membrane potential $\xi$ reaches its threshold $\Delta$.
Jump set $\mathcal{D}_2$ in \eqref{eq:jumpsetD2} shows that when $\xi$ reaches $\Delta$, the jump will occur at most $T_c$ time units later than $t_{\Delta, i+n}$,
which we write as $\beta T_c$, for some $0\leqslant \beta < 1$. 
We therefore have $t_{p(i)}-t_i = k_c T_c$, where $k_c \in \mathbb{Z}_{>0}$ represents the number of jumps of the solution $q$ between $t_i$ and $t_{p(i)}, i \in \{1,2, \ldots, j\}$. 
A visual representation is given in Fig.~\ref{fig:xi_overshoot}.

\begin{figure}[thpb]
    \centering
    \includegraphics[width=0.7\linewidth]{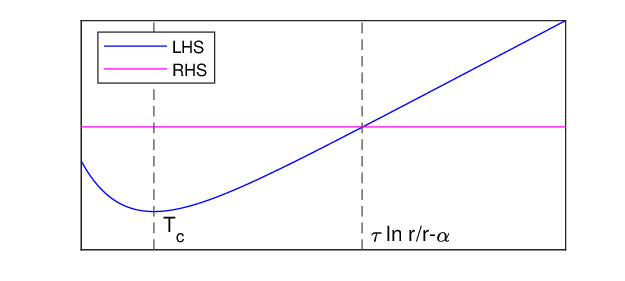}
     \caption{LHS and RHS of \eqref{eq:dwell_time_constant_RHS} in function of the time $\tau_{d}^i$.}
      \label{fig:dwell_time_shape}
\end{figure}

Thus, at $t_{\Delta, i+n}$, \eqref{eq:neuron_between_pellets} becomes
\begin{gather}
\begin{aligned}
    \xi&(t_{\Delta, i+n},i+n) \\
    &= r(t_{\Delta, i+n}-t_i) - \tau (r-x(t_i,i)) \left(1-e^{-\frac{1}{\tau}(t_{\Delta, i+n}-t_i)}\right) \\ 
    &= \Delta .
    \label{eq:xi_at_t_delta}
\end{aligned}
\raisetag{1\baselineskip}
\end{gather}
Using \eqref{eq:pellet_jump_time} and \eqref{eq:xi_at_t_delta},
and since $t_{\Delta, i+n} = t_{p(i)}-\beta T_c$, we get for each
$i \in \{1,2,\ldots,j\}$ such that $(t_i,i-1) \in \mathcal{P}(q)$,
\begin{multline}
t_{p(i)}-t_i = \\
\frac{1}{r}\left(\Delta+\tau
    (r-x(t_i,i)) \left(1-e^{-\frac{1}{\tau}(t_{p(i)}-t_i-\beta T_c)}\right) \right)+\beta T_c.
    \label{eq:dwell_time}
\raisetag{2.6\baselineskip}
\end{multline}
Using $x(t_i,i) \in (0,r]$ and $\beta \in [0,1)$, we get
\begin{equation}
t_{p(i)}-t_i \leqslant \frac{\Delta}{r}+\tau
     \left(1-e^{-\frac{1}{\tau}(t_{p(i)}-t_i-T_c)}\right) + T_c.
     \label{eq:dwelltime_with_delta}
\end{equation}
Given that
$\Delta \leqslant r \tau \ln{\frac{r}{r-\alpha}} - r \tau \left(1- \frac{r-\alpha}{r} e^{\frac{T_c}{\tau}}\right) - r T_c$, by \eqref{eq:Delta_condition}, 
and note that $r \tau \ln{\frac{r}{r-\alpha}} - r \tau \left(1- \frac{r-\alpha}{r} e^{\frac{T_c}{\tau}}\right) - r T_c>0$ since $T_c \in \left(0, \tau \ln \frac{r}{r-\alpha}\right]$ from \eqref{eq:Tc_condition},
\eqref{eq:dwelltime_with_delta} becomes
\begin{equation}
t_{p(i)}-t_i \leqslant \tau \ln{\frac{r}{r-\alpha}} + \tau \frac{r-\alpha}{r} e^{\frac{T_c}{\tau}}-\tau
     e^{-\frac{1}{\tau}(t_{p(i)}-t_i)} e^{\frac{T_c}{\tau}} .
\end{equation}
Rearranging the terms, gives 
\begin{equation}
t_{p(i)}-t_i +\tau e^{-\frac{1}{\tau}(t_{p(i)}-t_i)} e^{\frac{T_c}{\tau}} \leqslant \tau \ln{\frac{r}{r-\alpha}} + \tau \frac{r-\alpha}{r} e^{\frac{T_c}{\tau}}.
\label{eq:dwell_time_constant_RHS}
\end{equation}
To find the maximum time $t_{p(i)} - t_i$ between two pellet launches, we evaluate the left-hand side (LHS) as a function of the time $\tau_{d}^i:=t_{p(i)}-t_i$, and compare it to the right-hand side (RHS) of \eqref{eq:dwell_time_constant_RHS}, which is constant.
The shape of the LHS compared to a constant RHS is illustrated in Fig.~\ref{fig:dwell_time_shape}.
Recall that the minimum dwell-time is determined by the actuator (related to timer $T$ in the model), so  at least it holds that
$   \tau_{d}^i \geqslant T_c$.
Taking the derivative of the LHS of \eqref{eq:dwell_time_constant_RHS}, we get
\begin{equation}
    \frac{d}{d\tau_{d}^i} \left(\tau_{d}^i +\tau e^{-\frac{1}{\tau}\tau_{d}^i} e^{\frac{T_c}{\tau}} \right) 
    =1-e^{\frac{T_c}{\tau}}e^{-\frac{\tau_{d}^i}{\tau}}.
    \label{eq:dwelltime_derivative}
\end{equation}
We can verify that \eqref{eq:dwelltime_derivative} has a zero at $\tau_{d}^i = T_c$, and is positive for all $\tau_{d}^i \geqslant T_c$, so the LHS of \eqref{eq:dwell_time_constant_RHS} is monotonically increasing for $\tau_{d}^i \geqslant T_c$. 
Also, since the equality in \eqref{eq:dwell_time_constant_RHS} holds at $\tau_{d}^i = \tau \ln{\frac{r}{r-\alpha}}$, 
the maximum time between two pellets for which \eqref{eq:dwell_time_constant_RHS} holds when $x(t_i,i-1)>0$, 
for all $i \in \{1,2,\ldots,j\}$, can be simplified to
\begin{equation}
t_{p(i)}-t_i = \tau_{d}^i \leqslant \tau \ln \frac{r}{r-\alpha}=:\tau_d.
\label{eq:max_dwell_time}
\end{equation}
Note that $i=0$ is not considered above since there is no pellet launch at $t_0$, given the initial conditions.
However, recalling that $t_0=0$ and from $T(0,0)=\xi(0,0)=0$, 
a similar logic follows also for $i=0$, and thus
\eqref{eq:xi_at_t_delta}-\eqref{eq:max_dwell_time} hold for $i\in \{0,1,\ldots,j\}$.

\noindent \textbf{Decrease of $x$ when a pellet is launched}\\
For $i \in \{0,1,\ldots,j\}$, we now consider the decrease in $x$ for a jump due to a pellet launch in $x(t_{p(i)}, i+n) = x(t_{p(i)}, p(i) -1) \in \mathcal{D}_2$.
From \eqref{eq:jump_x} and \eqref{eq:pellet_jump_time}, we have,
\begin{equation}
    x(t_{p(i)},p(i)) = x(t_{p(i)},i+n)-\alpha,
    \label{eq:V_decrease_pellet}
\end{equation}
where we recall that $t_{p(i)}$ is the first time after $t_i$ at which a pellet is fired,
and that, between $t_i$ and $t_{p(i)}$ there are $n \in \mathbb{Z}_{\geqslant 0}$ jumps generated by the timer, where no pellet is fired.

\noindent \textbf{Guaranteeing overall decrease}\\
Given the increase in $x$ in \eqref{eq:V_increase_between_pellets}, the maximum time between two consecutive pellet jumps in \eqref{eq:max_dwell_time}, and the decrease in $x$ in \eqref{eq:V_decrease_pellet},
we now show that there is a net decrease of $x$ over one cycle, consisting of increase between 2 consecutive pellet launches, and decrease in one pellet launch. 
At $t_{p(i)}$, with $p(i)$ defined in \eqref{eq:pellet_jump_time} for each 
$i \in \{0,1, \dots, j\}$, from  \eqref{eq:V_decrease_pellet} and \eqref{eq:V_increase_between_pellets}, we get
\begin{equation}
    x(t_{p(i)},p(i)) = r-e^{-\frac{1}{\tau}(t_{p(i)}-t_i)}\left(r-x(t_i,i)\right) -\alpha .
    \label{eq:x_flow_and_pelletjump}
\end{equation}
Here we recall that the cycle starts right after a previous pellet launch at $(t_i,i)$ 
for all $i \in \{1,2\ldots,j\}$ such that $(t_i,i-1) \in \mathcal{P}(q)$, 
or from the initial conditions $T(0,0)=\xi(0,0)=0$ if $i=0$,
with $x(t_i,i) >0$ the error at the beginning of the cycle, and $x(t_{p(i)},p(i))$ the error at the end, after the next pellet launch. 
Since $x(t_i,i) \in (0,r]$, and using \eqref{eq:max_dwell_time}, \eqref{eq:x_flow_and_pelletjump} becomes
\begin{gather}
\begin{split}
    x(&t_{p(i)},p(i)) \\
    =& \Bigg(\frac{r\left(1-e^{-\frac{1}{\tau}(t_{p(i)}-t_i)}\right)-\alpha}{x(t_i,i)}
    +e^{-\frac{1}{\tau}(t_{p(i)}-t_i)}\Bigg)
    x(t_i,i) \\
    \leq& \left(\frac{r\left(1-\frac{r-\alpha}{r}\right)-\alpha}{x(t_i,i)}
    +\frac{r-\alpha}{r}\right) x(t_i,i) \\
    =& \; \frac{r-\alpha}{r} \; x(t_i,i) \\
    =& \; \gamma x(t_i,i) ,
\end{split}
\label{eq:x_decrease_flow_and_jump}
\raisetag{1\baselineskip}
\end{gather}
with $\gamma = \frac{r-\alpha}{r} < 1$.
Therefore, for all $i \in \{0,1,\ldots,j\}$ such that $x(t_i, i) >0$, 
$    x(t_{p(i)},p(i)) \leqslant \gamma x(t_i,i)$,
meaning that after a full cycle, including the continuous increase starting immediately after the last pellet jump in $(t_i,i)$, 
or at the initial condition $(t_0,0)$ when $i=0$,
and including the pellet jump in $(t_{p(i)}, p(i))$, $x$ has net decreased. 
Using the jump map in \eqref{eq:jump_map}-\eqref{eq:jump_map_G2}, and recalling that the jump size is equal to $\alpha$, we have that for all 
$s \in [t_i,t_{p(i)}]$ such that $x(t_i,i) \in (0,r]$,
\begin{equation}
    -\alpha < x(s,\cdot) \leqslant \gamma x(t_i,i) + \alpha .
    \label{eq:stabilityOneFlowOneJump}
\end{equation}
Pick any $(t^\star, j^\star-1), (t,j) \in \text{dom } q$ such that  $(t^\star, j^\star-1) \in \mathcal{P}(q) \cup (0,0)$,  $(t^\star,j^\star-1) \leqslant (t,j)$ and $x(t^\star, j^\star-1)>0$. 
Denote $n(t,t^\star)$ the number of pellet launches that occur between $(t^\star,j^\star-1)$ and $(t,j)$, using \eqref{eq:max_dwell_time} we have that 
\begin{equation}
    t-t^\star \leqslant (n(t,t^\star) +1) \tau_d, 
    \label{eq:MaxDwellTimeTot}
\end{equation}
which corresponds to 
\begin{equation}
    n(t,t^\star) \geqslant \frac{t-t^\star}{\tau_d} -1. 
    \label{eq:maximumAverageDwellTime}
\end{equation}
Moreover, using \eqref{eq:stabilityOneFlowOneJump} we have that, for all $s \in (t^\star,t)$,
\begin{equation}
    -\alpha < x(s,\cdot) \leqslant \gamma^{n(t,t^\star)} x(t^\star,j^\star) + \alpha ,
    \label{eq:stabilityOneFlowOneJump2}
\end{equation}
which, since $\gamma \in (0,1)$ and using \eqref{eq:maximumAverageDwellTime}, becomes
\begin{equation}
    -\alpha < x(s,\cdot) \leqslant \gamma^{\frac{t-t^\star}{\tau_d} -1} x(t^\star,j^\star) + \alpha ,
\end{equation}
Consequently, we have that for all 
$(t,j) \in \text{dom }q$ such that $x(0,0)>0$
\begin{equation}
    -\alpha < x(t,j) \leqslant \gamma^{\frac{t}{\tau_d} -1} x(0,0) + \alpha .
    \label{eq:xPosFinal}
\end{equation}
Moreover, since the solution $q$ is $t$-complete in view of Proposition~\ref{prop:hybrid-model-basic-properties}, from \eqref{eq:xPosFinal} we have that, for any $x(0,0) >0$, any corresponding maximal solution to \eqref{eq:hybrid_system_general} satisfies 
$\lim_{t+j \rightarrow + \infty} \sup |x(t,j)| \leqslant \alpha$. 

\noindent \textbf{In case $x(t_i,i) \leqslant 0$}\\
Until now, we have considered only $x(t_i,i)>0$, we now look at the case where $x(t_i,i) \leqslant 0$ for all $i \in \{0,1,\dots, j\}$.
From the neuron dynamics in \eqref{eq:neuron_flow}, no control actions occur when $x \leqslant 0$, as $\max{(0,x)} = 0$ in \eqref{eq:neuron_flow}. 
Since \eqref{eq:openloop_x} shows that $x$ will always increase without control actions the controller will ``wait'' until it starts integrating again when $x>0$.
We define for each $i \in \{0,1, \dots, j\}$, such that $x(t_i,i)\leqslant 0$,
$\bar t:= \inf\{s\geqslant t_i: (s,i) \in \textnormal{dom } q \textnormal{ and } x(s,i) = 0\}$.
Therefore, in view of \eqref{eq:openloop_x}, we have $x(\underline{s},i) \leqslant 0$ for all $\underline{s} \in [t_i,\bar t]$ and $x(\bar s,i) > 0$ for all $\bar s \in (\bar t, t_{p(i)}]$, with $p(i)$ defined in \eqref{eq:pellet_jump_time}.
Then, using similar steps as in the case where $x(t_i,i)> 0$, we have
\begin{equation}
    -\alpha < x(\bar s,\cdot) \leqslant \alpha.
    \label{eq:xNeg1}
\end{equation}
Moreover, for all $\underline{s} \in [t_i,\bar t]$, 
 from \eqref{eq:openloop_x}, we have 
\begin{equation}
    x(\underline{s},\cdot) = r-e^{-\frac{\underline{s}-t_i}{\tau}}\left(r-x(t_i,i)\right).
    \label{eq:xNegFinal1}
\end{equation} 
Consequently, from \eqref{eq:xNeg1} and \eqref{eq:xNegFinal1}, we have that for all 
$(t,j) \in \text{dom }q$, such that with $x(0,0) \leqslant 0$:
\begin{align}
   \text{min}\left(r-e^{-\frac{t}{\tau}}\left(r-x(0,0)\right),-\alpha\right) < x(t,j) \leqslant  \alpha .
   \label{eq:xNegFinal2}
\end{align}
Moreover, merging \eqref{eq:xPosFinal} and \eqref{eq:xNegFinal2} and since any maximal solution $q$ is $t$-complete in view of Proposition~\ref{prop:hybrid-model-basic-properties}, we have that, for any $x(0,0) \leqslant r$, 
any corresponding maximal solution to \eqref{eq:hybrid_system_general} satisfies 
$\lim_{t+j \rightarrow +\infty} \sup |x(t,j)| \leqslant \alpha$. 
This concludes the proof.\hfill $\blacksquare$

\section{Proof of Proposition~\ref{prop:inputclipping}}
\label{app:inputclipping}
The proof of Proposition~\ref{prop:inputclipping} follow similar lines as the proof of Theorem~\ref{theorem:main}.
We therefore explicitly provide the main steps, highlighting the differences with respect to the Proof of Theorem~\ref{theorem:main}.

Consider system $\mathcal{H}_{\textnormal{IC}}$ in \eqref{eq:hybrid_system_ic}.
Compared to the proof of Theorem~\ref{theorem:main}, we now separate the dynamics of $\xi$ during flow in \eqref{eq:flow_with_saturation}:
\begin{enumerate}
\item[(i)] $\dot{\xi} = 0  \quad \text{when} \quad    x < 0 $
\item[(ii)] $\dot{\xi} = x  \quad \text{when} \quad    0 \leqslant x < \frac{\Delta}{T_c}$
\item[(iii)] $\dot{\xi} = \frac{\Delta}{T_c}  \quad \text{when} \quad  \frac{\Delta}{T_c}  \leqslant x$
\end{enumerate}
Given that
$\Delta \leqslant \left(r - (r-\alpha) e^{\frac{2 T_c}{\tau}}\right)T_c$
by \eqref{eq:Delta_condition_SDM}, 
and note that $\left(r - (r-\alpha) e^{\frac{2 T_c}{\tau}}\right)T_c>0$
since $T_c \in \left(0, \frac{\tau}{2} \ln \frac{r}{r-\alpha}\right]$ from \eqref{eq:Tc_condition_SDM}, we have $\frac{\Delta}{T_c} > 0$.
We consider these three cases separately.

\noindent\emph{(i)} The absence of control actions when $x<0$ establishes the lower bound \eqref{eq:ultimate_bound_input_clipping} of Proposition~\ref{prop:inputclipping}.
Similar to the proof of Theorem~\ref{theorem:main}, we consider the case where $x(t_i,i) \leqslant 0$.
From the neuron dynamics in \eqref{eq:flow_with_saturation}, we have that no control actions occur when $x \leqslant 0$, as $\dot \xi = 0$ in \eqref{eq:flow_with_saturation}. 
Since \eqref{eq:openloop_x} shows that $x$ will always increase without control actions the controller will ``wait'' until it starts integrating again. 
Similar to the case where $x(t_i,i)\leqslant0$ in the proof of Theorem~\ref{theorem:main}, we again define for each $i \in \{0,1, \dots, j\}$, such that $x(t_i,i)\leqslant 0$,
$\bar t:= \inf\{s\geqslant t_i: (s,i) \in \textnormal{dom } q \textnormal{ and } x(s,i) = 0\}$.
Therefore, we have $x(\bar s,i) > 0$ for all $\bar s \in (\bar t, t_{p(i)}]$, with $p(i)$ defined in \eqref{eq:pellet_jump_time}.
Using the jump map in \eqref{eq:jump_map_sigmadelta}-\eqref{eq:jump_map_G2_sigmadelta}, and recalling that the jump size is equal to $\alpha$, we have, similar to \eqref{eq:xNeg1}, that for all $\bar s \in (\bar t,t_{p_i}]$
\begin{equation}
    -\alpha < x(\bar s,\cdot).
\end{equation}
For all $\underline{s} \in [t_i,\bar t]$, \eqref{eq:xNegFinal1} holds.
Consequently, we have that for all 
$(t,j) \in \text{dom }q$,
such that
with $x(0,0) \leqslant 0$:
\begin{align}
   \text{min}\left(r-e^{-\frac{t}{\tau}}\left(r-x(0,0)\right),-\alpha\right) < x(t,j).
   \label{eq:xNegFinal2SDM}
\end{align}
Note that, although the method follows the same steps as the proof in Appendix~\ref{app:main}, \eqref{eq:xNeg1}-\eqref{eq:xNegFinal2}, we can only derive a lower bound here.

\noindent\emph{(ii)} 
Let $0 \leqslant x(t_i,i) < \frac{\Delta}{T_c}$.
Define $t_{\text{sat}, i} := \inf\{s\geqslant t_{i} : x(s, i) = \frac{\Delta}{T_c}\}$ as the time $x$ reaches its saturation point $\frac{\Delta}{T_c}$, 
We first assume that this saturation point exists, i.e., $t_{\text{sat}, i} \in [t_i,t_{i+1}]$, and later we will consider the case when it does not exist.
From the dynamics of $\xi$ in \eqref{eq:flow_with_saturation}, we get
\begin{align}
    \xi(t_{i+1},i) =& \xi(t_i,i) + \int_{t_i}^{t_{i+1}} \text{sat}(x(\tilde{s},i)) d\tilde{s}
     \label{eq:xi_case2_sat}\\
      =& \xi(t_i,i) + \int_{t_i}^{t_\text{sat}} x(\tilde{s},i) d\tilde{s}+\int_{t_\text{sat}}^{t_{i+1}}\frac{\Delta}{T_c} d\tilde{s}.
      \hspace*{27pt}
    \label{eq:xi_case2_full}
\raisetag{1.4\baselineskip}    
\end{align}
Since $x(\tilde{s},i) < \frac{\Delta}{T_c}$ for all $\tilde{s} \in [t_i,t_\text{sat}]$, and $t_{i+1}-t_i=T_c$, we get
\begin{align}
    \int_{t_i}^{t_\text{sat}} x(\tilde{s},i) d\tilde{s}+\int_{t_\text{sat}}^{t_{i+1}}\frac{\Delta}{T_c} d\tilde{s} < \Delta.
    \label{eq:xi_case2_integrals}
\end{align}
Note that $\xi(t_i,i) \in [0,\Delta)$, and since a pellet is triggered at $\xi(t_{i+1},i)\geqslant \Delta$, we can see from \eqref{eq:xi_case2_full} that we get the most conservative case (where the most time is needed to trigger a pellet) for $\xi(t_i,i) = 0$.
Indeed, any $\xi(t_i,i) > 0$ will result in $\xi$ reaching $\Delta$ sooner.
Considering this conservative case with $\xi(t_i,i) = 0$, we get from \eqref{eq:xi_case2_full} and \eqref{eq:xi_case2_integrals} that
\begin{equation}
    \xi(t_{i+1},i) < \Delta.
    \label{eq:xi_case2_conservative}
\end{equation}
Note that if $t_{\text{sat}, i}$ does not exist, i.e., the saturation point is not reached before the next jump at $(t_{i+1},i)$, we can simplify the integrals in \eqref{eq:xi_case2_sat}-\eqref{eq:xi_case2_integrals} and still get the same result as in \eqref{eq:xi_case2_conservative}.
More general, this means that for $0 \leqslant x(t_i,i) < \frac{\Delta}{T_c}$, a scenario exists (depending on the value of $\xi(t_i,i)$) with no pellet launch at time $t_{i+1}$.

\noindent\emph{(iii)}
Let $x(t_i,i) \geqslant \frac{\Delta}{T_c}$,
for each $i \in \{0,1,\ldots,j\}$ and all $s \in [t_i,t_{i+1}]$,
from the neuron dynamics \eqref{eq:flow_with_saturation}, we have
    $\xi(s,i) = \xi(t_i,i) + \frac{\Delta}{T_c}(s-t_i)$.
The increase of $\xi$ between two jumps, when $x(t_i,i) \geqslant \frac{\Delta}{T_c}$, is exactly
    $\xi(t_{i+1},i) = \xi(t_i,i) + \frac{\Delta}{T_c}(t_{i+1}-t_i)= \xi(t_i,i) +\Delta \geqslant \Delta$,
where we used $t_{i+1}- t_i = T_c$ from the timer dynamics \eqref{eq:timer_flow}-\eqref{eq:reset_timer}.
Hence, $(t_{i+1},i) \in \mathcal{P}(q)$, and a pellet is fired.
Using \eqref{eq:reset_neuron_SDM}, we get
    $\xi(t_{i+1},i+1) = \xi(t_{i+1},i) - \Delta = \xi(t_i,i)$.
So, for all $x(t_i,i) \geqslant \frac{\Delta}{T_c}$, a pellet is shot at the next possible launch slot, and $t_{p(i)}-t_i=t_{i+1}-t_i=T_c$.
Since, from the condition in \eqref{eq:Tc_condition_SDM}, we have $T_c\leqslant\frac{\tau}{2}\ln{\frac{r}{r-\alpha}} <\tau\ln{\frac{r}{r-\alpha}}$, we can guarantee the combined increase of $x$ between two pellet jumps, and pellet jumps, decreases over time, similar to Theorem \ref{theorem:main}.

Combining (ii) and (iii), we can determine the upper bound on $x$.
For all $i \in \{0,1,\dots, j\}$, let $x(t_i,i) = \lim_{\eta \rightarrow 0} \left( \frac{\Delta}{T_c}-\eta \right)$ and consider again the most conservative case with $\xi(t_i,i)=0$.
At hybrid time $(t_i,i)$, since $0 \leqslant x(t_i,i) < \frac{\Delta}{T_c}$, we are in case (ii).
Given the dynamics of $\xi$ in \eqref{eq:xi_case2_sat}-\eqref{eq:xi_case2_conservative}, there will not be a pellet launch at the next jump time $t_{i+1}$.
However, at hybrid time $(t_{i+1},i+1)$, for $x$ we have $\frac{\Delta}{T_c} \leqslant x(t_{i+1},i+1)$.
Hence we continue the next flow in case (iii),
and there will be a pellet launch at the next jump time $t_{i+2}$.
So, to determine the upper bound, we have to calculate the increase of $x$ during 2 flows, starting right below the saturation point where $x(t_i,i) = \lim_{\eta \rightarrow 0} \left( \frac{\Delta}{T_c}-\eta \right)$ and using $t_{p(i)}-t_i = t_{i+2}-t_i = 2T_c$.
From \eqref{eq:V_increase_between_pellets}, we calculate the upper bound of the steady-state error $x_{\text{UB}}$
\begin{align}
    x_{\text{UB}}&=x(t_{p(i)},p(i)-1) \\
        &= r-e^{-\frac{t_{p(i)}-t_i}{\tau}}\left(r-x(t_i,i)\right)\\
        &< r-e^{-\frac{2T_c}{\tau}}\left(r-\frac{\Delta}{T_c}\right)
\end{align}
Using the constraint on $\Delta$ in \eqref{eq:Delta_condition_SDM}, we get
\begin{align}
    x_{\text{UB}}&< r-e^{-\frac{2T_c}{\tau}}+\left(\alpha-r+re^{-\frac{2T_c}{\tau}}\right) < \alpha.
    \label{eq:SDM_UB}
\end{align}
Combining \eqref{eq:xNegFinal2SDM} with \eqref{eq:SDM_UB}, we get the bounds in \eqref{eq:ultimate_bound_input_clipping}.
This concludes the Proof of Proposition~\ref{prop:inputclipping}.\hfill $\blacksquare$

\begin{rem}
\label{rem:SDM_same_speed_NM}
When the actuator speed is constrained in a similar way as the neuromorphic controller, i.e., $T_c \in (0,\tau \ln{\frac{r}{r-\alpha}}]$, we can calculate the upper bound $x_\text{UB}$ as above.
For $\lim_{\Delta \rightarrow 0}$, which is the fastest controller, we can guarantee $x_\text{UB} \leqslant 2\alpha-\frac{\alpha^2}{r} = \alpha(2-\frac{\alpha}{r})  <2\alpha$, since $r>\alpha$.
\end{rem}

\section{Proof of Proposition~\ref{prop:jumpmap}}
\label{app:sdm2}
To prove Proposition~\ref{prop:jumpmap}, we can follow similar steps as in the proof of Theorem~\ref{theorem:main} in Appendix~\ref{app:main}.
Compared to Theorem~\ref{theorem:main} where, after a pellet launch, the neuron membrane potential $\xi(t_i,i) = 0$, in Proposition~\ref{prop:jumpmap}, after a pellet launch, the neuron membrane potential $\xi(t_i,i) \in [0,\Delta)$ in view of \eqref{eq:jump_map_sigmadelta2}-\eqref{eq:jump_map_G2_sigmadelta2}.
Thus, \eqref{eq:xi_total_flow} becomes
$    \xi(s,\cdot) = \xi(t_{i},i) + \int_{t_i}^{s} x(\tilde{s},\cdot)d\tilde{s}$.
This increases the likelihood that the neuron membrane potential reaches its threshold $\Delta$ sooner, resulting in less time between two pellets.
So we expect this controller to be faster in general, however, this is not a guarantee, 
as there could exist jump times $(t_i,i) \in \text{dom }q$ such that $\xi(t_i,i) = 0$. 
Therefore, the limit case remains the same, and we have to use $t_{p(i)}-t_i$ as calculated in \eqref{eq:dwell_time} and get the same stability property as before.\hfill $\blacksquare$

\section{Proof of Proposition~\ref{prop:refractory}}
\label{app:refr}
To prove Proposition~\ref{prop:refractory}, consider system $\mathcal{H}_{\textnormal{PP}}$ in \eqref{eq:hybrid_system_PP} and define 
\begin{equation}
    \overline{T}_c:=lT_c,
    \label{eq:def_T_c_bar}
\end{equation}
with $l= \Bigl \lceil \frac{T_{\textnormal{prep}}}{T_c} \Bigr \rceil \in \mathbb{N}$.
Now adjust the jump set in \eqref{eq:jumpset_PP}-\eqref{eq:jumpsetD2_PP} and the jump map in \eqref{eq:jump_map_PP}-\eqref{eq:jump_map_G2_PP}, replacing $T_c$ with $\overline T_c$.
This new system, which we will call $\overline{\mathcal{H}}_{\text{PP}}$, is a hybrid system with similar dynamics but potentially a slower actuator (if $l>1$), compared to $\mathcal{H}_{\text{PP}}$.
Hence, the stability analysis of $\overline{\mathcal{H}}_{\text{PP}}$ is more conservative, since the time between two jumps is longer than in system $\mathcal{H}_{\textnormal{PP}}$ in \eqref{eq:hybrid_system_PP}.

Since $\overline{T}_c \geqslant T_{\text{prep}}$, we have $T_p\geqslant T_{\text{prep}}$ at every jump in $\overline{\mathcal{H}}_{\text{PP}}$.
Indeed, the pellet preparation period is over before the actuator timer reaches its threshold.
As a result, $T_p$ will not be decisive in determining if $\tilde q \in \mathcal{D}_\textnormal{PP1}$ or $\tilde q \in \mathcal{D}_\textnormal{PP2}$ at a jump time in \eqref{eq:jump_map_PP}-\eqref{eq:jump_map_G2_PP}, this is only determined by $T$ and $\xi$.
Therefore, $T_p$ can be ignored in model \eqref{eq:hybrid_system_PP},
it will not introduce any additional delays in firing a pellet. 
Hence, we can apply Theorem~\ref{theorem:main}
in the NM case where $\varepsilon = \xi$,
or Proposition~\ref{prop:jumpmap} in the SDM case where $\varepsilon = k\Delta$,
to a system with $\overline{T}_c$ as synchronisation time.
Using \eqref{eq:Tc_condition} and \eqref{eq:Delta_condition} from Theorem~\ref{theorem:main}, or \eqref{eq:Tc_condition_JM} and \eqref{eq:Delta_condition_JM} from Proposition~\ref{prop:jumpmap}, leads in both cases to the conditions 
\begin{equation}
    \overline T_c \in \left(0,\tau \ln{\frac{r}{r-\alpha}}\right],    
    \label{eq:Tc_condition_accent}
\end{equation}
and 
\begin{align}
\Delta \in \left[0,r \tau \ln{\frac{r}{r-\alpha}} - r \tau \left(1- \frac{r-\alpha}{r} e^{\frac{\overline T_c}{\tau}}\right) - r \overline T_c\right].
\label{eq:Delta_condition_accent}
\hspace*{20pt}
\raisetag{1.4\baselineskip} 
\end{align}
Given \eqref{eq:Tprep_condition_PP} and \eqref{eq:def_T_c_bar}, we can write \eqref{eq:Tc_condition_accent}-\eqref{eq:Delta_condition_accent} as \eqref{eq:Tc_condition_PP}-\eqref{eq:Delta_condition_PP}.
This concludes the proof of Proposition~\ref{prop:refractory}.\hfill $\blacksquare$